\documentclass[pdflatex,sn-mathphys-num]{sn-jnl}

\usepackage{graphicx}%
\usepackage{multirow}%
\usepackage{amsmath,amssymb,amsfonts}%
\usepackage{amsthm}%
\usepackage{mathrsfs}%
\usepackage[title]{appendix}%
\usepackage{xcolor}%
\usepackage{textcomp}%
\usepackage{manyfoot}%
\usepackage{booktabs}%
\usepackage{algorithm}%
\usepackage{algorithmicx}%
\usepackage{algpseudocode}%
\usepackage{listings}%
\usepackage{physics}%
\usepackage{cleveref}%
\usepackage{subcaption}%

\theoremstyle{thmstyleone}%
\theoremstyle{thmstyletwo}%

\theoremstyle{thmstylethree}%

\begin{document}

\title[Photon Fragmentation Functions in \textsc{LHAPDF}: Existing Sets and a \textsc{Pythia}-Based Extraction]{Photon Fragmentation Functions in \textsc{LHAPDF}: Existing Sets and a \textsc{Pythia}-Based Extraction}


\author*[1,2]{\fnm{Alexander Puck} \sur{Neuwirth}}\email{alexander.neuwirth@unimib.it}

\affil*[1]{\orgdiv{Dipartimento di Fisica “Giuseppe Occhialini”}, \orgname{Universit\`a degli Studi di Milano-Bicocca}, \orgaddress{\street{Piazza della Scienza 3}, \city{Milano}, \postcode{20126}, \state{Lombardy}, \country{Italy}}}

\affil[2]{\orgdiv{Sezione di Milano-Bicocca}, \orgname{INFN}, \orgaddress{\street{Piazza della Scienza 3}, \city{Milano}, \postcode{20126}, \state{Lombardy}, \country{Italy}}}


\abstract{
Prompt-photon predictions with experimentally relevant isolation criteria retain a residual dependence on parton-to-photon fragmentation functions, whose phenomenological uncertainties remain only weakly constrained.
We present a pragmatic study of photon fragmentation functions extracted from \textsc{Pythia} parton-shower events and stored in \textsc{LHAPDF} grids.
We briefly review the role of photon fragmentation in perturbative QCD and then construct shower-level estimates of the flavour-dependent functions $D^{\mathrm{PS}}_{\gamma/i}(z,\mu_D)$ as functions of the photon momentum fraction $z$ and fragmentation scale $\mu_D$.
The extracted distributions show the expected qualitative hierarchy between quark and gluon channels and an increasing fragmentation contribution with scale.
In parallel, we convert the commonly used BFG, GRV, and GDRG parameterizations into \textsc{LHAPDF} grids and study their interpolation.
We find that the default \texttt{logcubic} interpolation in \textsc{LHAPDF} reproduces the non-trivial $z$- and scale-dependence of the tabulated fragmentation functions more faithfully than the linear interpolation commonly used in legacy implementations.
We also define a simple collider-oriented envelope set, based on the BFG and GRV families. These results provide a portable baseline for future phenomenological studies and for more systematic extractions of photon fragmentation functions.
}




\maketitle

\section{Introduction}
\label{sec:intro}

Prompt photons are a classic probe of perturbative quantum chromodynamics (QCD) and remain phenomenologically important at modern colliders.
They test short-distance quark--gluon dynamics, constrain parton densities, and form an irreducible background to a variety of Standard-Model (SM) and beyond-the-Standard-Model (BSM) measurements \cite{Klasen:2002xb,Catani:2002ny,Aurenche:2006vj,dEnterria:2012kvo}.
Their theoretical description contains both a direct component, in which the photon participates in the hard scattering, and a fragmentation component, in which a high-transverse-momentum ($p_T$) parton produces a photon during final-state collinear radiation \cite{Catani:2002ny,Aurenche:2006vj}.
While photon isolation suppresses the latter contribution, realistic collider definitions do not remove it completely, so precision predictions still depend on parton-to-photon fragmentation functions (FFs) \cite{Frixione:1998hn,Siegert:2016bre,Chen:2019zmr,Chen:2022gpk,Gehrmann:2022cih}.

Compared with parton distribution functions (PDFs) and hadron fragmentation functions~\cite{Soleymaninia:2026edd}, photon fragmentation functions are only weakly constrained.
The widely used GRV (Gl\"uck--Reya--Vogt) \cite{Gluck:1992zx} and BFG (Bourhis--Fontannaz--Guillet) \cite{Bourhis:1997yu} parameterizations are built from perturbative evolution supplemented by non-perturbative input, commonly modeled through vector-meson dominance (VMD), and are anchored mainly by measurements at the Large Electron--Positron Collider (LEP) \cite{OPAL:1997lep,Gehrmann-DeRidder:1998bju}.
Subsequent studies showed that low-$p_T$ proton--proton ($pp$) data, photon--jet correlations, and photon-in-jet observables can provide additional sensitivity, but the available constraints remain limited \cite{Belghobsi:2009hx,Klasen:2014xfa,Kaufmann:2016nux,Kaufmann:2017lsd,Kaufmann:2017jpb}.
This has become more relevant as next-to-next-to-leading-order (NNLO) calculations with realistic photon isolation have turned the treatment of fragmentation into a practical phenomenological limitation rather than only a formal one \cite{Chen:2022gpk,Gehrmann:2022cih}.
Very recently, the sPHENIX Collaboration also reported an isolated prompt-photon measurement in $pp$ collisions at $\sqrt{s}=200$ GeV comparing to NLO and NNLO predictions with BFG and GRV fragmentation functions and \textsc{Pythia} predictions \cite{sPHENIX:2026zet}.

The goal of this paper is to explore a pragmatic alternative source of photon fragmentation functions based on \textsc{Pythia}~\cite{Sjostrand:2006za,Sjostrand:2007gs,Sjostrand:2014zea,Bierlich:2022pfr}.
We study how such functions can be extracted from parton-shower events, assess whether their scale dependence reproduces the expected qualitative behaviour, and provide them in an \textsc{LHAPDF} grid \cite{Buckley:2014ana} that can be used by external phenomenology tools.
This is intended as a flexible baseline for prompt-photon studies and as a complement to the traditional GRV- and BFG-type parameterizations.
In next-to-leading-order (NLO) computations of prompt photon production, the fragmentation contribution has been implemented via \textsc{Pythia} in the \textsc{POWHEG\,BOX} framework \cite{Nason:2004rx,Frixione:2007vw,Alioli:2010xd} instead of using the traditional fragmentation functions \cite{Jezo:2016ypn,Jezo:2024wsc}.
This work is also in line with Ref.~\cite{Jezo:2025jyu}, while that study focused on $\gamma \to e^+e^-$ conversions in the Kroll--Wada and parton-shower approaches, here we instead study jet $\to\gamma$ fragmentation in \textsc{Pythia} and compare the extracted fragmentation functions with others.
Using the parton-shower approach has also been shown to work in \textsc{Sherpa}, where an interleaved QCD+QED shower supplemented with matrix-element merging was found to reproduce the ALEPH measurement of the photon fragmentation function in hadronic $Z^0$ decays \cite{Hoeche:2009xc,ALEPH:1995zdi}.

\textsc{LHAPDF} has recently also been extended to support neutrino structure functions \cite{Candido:2023utz} proving the versatility of the framework, hence it is natural to also provide photon fragmentation functions in a portable \textsc{LHAPDF} grid format.
Prompt photon production has also played a role in electron-proton collisions at HERA where the GRV photon fragmentation functions were studied in the context of the ZEUS experiment \cite{ZEUS:2006uxx,Chekanov:2007cf,Gehrmann-DeRidder:1999dey}.
Therefore, prompt photons and their fragmentation function might see a revival with the future Electron-Ion Collider (EIC) \cite{Accardi:2012qut,AbdulKhalek:2021gbh,Rein:2024fns}.
Existing tools such as \textsc{EPhox} operate analogously to \textsc{JetPhox}~\cite{Catani:2002ny} and can be used to study photon production in electron-proton collisions \cite{Fontannaz:2001ek,Fontannaz:2001nq,Fontannaz:2003yn,Aurenche:2011wu} and will profit from the development of more unified photon fragmentation functions.

\section{Prompt Photon Production Framework}

Before turning to the \textsc{LHAPDF} implementation of existing photon fragmentation functions and to their extraction from \textsc{Pythia}, we first recall the theoretical framework relevant for prompt-photon production.
In particular, we summarize the factorized description of the fragmentation contribution and the role of photon-isolation prescriptions, which determine to what extent fragmentation effects survive in phenomenologically relevant collider observables.
This also serves to establish the notation used throughout the remainder of the paper.

\subsection{Fragmentation}
\label{sec:fragmentation}

At leading power, the inclusive production of an identified hadron or photon can be expressed in terms of initial-state parton distribution functions (PDFs), a short-distance partonic cross section, and a final-state fragmentation function \cite{Collins:1989gx}.
For a process $AB \to CX$, with $X$ denoting the undetected hadronic remnant, one may write schematically
\begin{equation}
    \begin{split}
    \dd\sigma_{AB \to C X} (p_A,p_B,p_C)
    =&
    \sum_{a,b,c}\int_0^1\dd{x_a} \dd{x_b} \dd{z}
    f_{a/A}(x_a,\mu_{F})f_{b/B}(x_b,\mu_{F})
    \\
    &\dd \hat \sigma_{ab \to cX}(x_ap_A,x_bp_B,\frac{p_C}{z},\mu_{R},\mu_{F},\mu_{D})
    D_{C/c}(z,\mu_{D})
    \,.
    \end{split}
\end{equation}
Here the sum runs over the intermediate initial- and final-state partons.
For the photon case, $C=\gamma$, the fragmentation contribution is described by the functions $D_{\gamma/c}(z,\mu_{D})$, which encode the probability density associated with a parton $c$ to produce a photon carrying a momentum fraction $z$ at the fragmentation scale $\mu_{D}$.

Unlike ordinary hadron fragmentation, photon fragmentation functions obey inhomogeneous time-like evolution equations because the photon couples directly to quarks.
At leading order (LO) the relevant quantum electrodynamics (QED) splitting kernels are
\begin{align}
	P_{\gamma q}(z) &= Q_q^2 \frac{1 + (1-z)^2}{z} \,,
	\\
	P_{\gamma g}(z) &= 0 \,.
\end{align}
The evolution can be written schematically as
\begin{equation}
	\frac{\dd D_{\gamma/i}(z,\mu_D^2)}{\dd \log \mu_D^2}
	=
	\frac{\alpha}{2\pi} P_{\gamma i}(z)
	+
	\frac{\alpha_s(\mu_D^2)}{2\pi}
	\sum_{j \in \{q,\bar q,g\}}
	P_{ji}(z)\otimes D_{\gamma/j}(z,\mu_D^2)
	\,,
\end{equation}
where the first term is the point-like contribution and the second term mixes the quark and gluon channels through the usual QCD time-like evolution \cite{Gluck:1992zx,Bourhis:1997yu}.
At lowest order this gives
\begin{align}
	\label{eq:evoff}
	D_{\gamma/q}(z,\mu_D) &= \frac{\alpha}{2\pi}  P_{\gamma q}(z) \log\left(\frac{\mu_D^2}{\mu_0^2}\right)  + D^\text{np}_{\gamma/q}(z,\mu_0)
	\,,\\
	D_{\gamma/g}(z,\mu_D) &= 0 + \mathcal{O}(\alpha\alpha_s)
	\,.
\end{align}
The boundary condition at the starting scale $\mu_0$ is non-perturbative.
In the standard vector-meson-dominance picture it is modeled as
\begin{equation}
	D^\text{had}_{\gamma,i}(z,\mu_0) = \sum_{V = \rho, \omega, \phi} \frac{4\pi \alpha}{f_V^2} D_{V/i}(z,\mu_0)
	\,.
\end{equation}
Different assumptions for this non-perturbative input are responsible for much of the spread between the available parameterizations and are one of the main reasons why additional phenomenological constraints are still valuable \cite{Gluck:1992zx,Bourhis:1997yu,Kaufmann:2017lsd}.

\subsection{Isolation}\label{sec:isolation}

Experimentally, prompt photons must be separated from the much larger background of photons produced in hadron decays.
This is achieved by imposing isolation requirements on the hadronic activity around the photon candidate.
The standard fixed-cone isolation used by experiments limits the hadronic transverse energy inside a cone of radius $R$ around the photon direction.
This definition is close to the measured object, but it retains a residual fragmentation contribution \cite{Catani:2002ny,Aurenche:2006vj}.
The smooth-cone prescription of Frixione instead suppresses hadronic radiation continuously as the cone shrinks, thereby removing collinear parton-photon configurations and eliminating the fragmentation contribution in perturbation theory \cite{Frixione:1998hn}.
Hybrid isolation combines an inner smooth cone with an outer fixed cone and is used in modern next-to-next-to-leading order (NNLO) calculations as a compromise between perturbative cleanliness and experimental realism \cite{Siegert:2016bre,Chen:2019zmr,Chen:2022gpk}.

These can all be described by a single equation, which is then used to calculate the hadronic activity around the photon and to compare against the isolation energy
\begin{equation}
    \sum_{i}^{\Delta R_{\gamma,i} \leq r} p_T^i \leq  p_T^\text{iso} \left(\frac{r}{\epsilon R}\right)^{2n}
   \quad\quad \forall r \leq \epsilon R\,,
   \label{eq:isolation}
\end{equation}
where $R$ is the desired cone radius parameter, $\Delta R_{\gamma,i} = \sqrt{ (\Delta y_{\gamma,i})^2 + (\Delta \phi_{\gamma,i})^2 }$ is the distance of the $i$-th particle to the photon, $p_T^i$ is the transverse momentum of the $i$-th particle, and $r$ is a scanning parameter.
Here, $(n=0,\epsilon=1)$ corresponds to the fixed-cone isolation, $(n=1,\epsilon=1)$ to the smooth-cone isolation, and the hybrid isolation requires both a smaller smooth-cone $(n=1,\epsilon=0.1)$ and a fixed cone $(n=0,\epsilon=1)$ to be satisfied \cite{Amoroso:2020lgh,Neuwirth:2024juw,Jezo:2024wsc}.
The form of \cref{eq:isolation} reveals a simple universal algorithmic implementation.
By starting iterating from the closest particle to the photon, moving outwards and updating $r$ to $\Delta R_{\gamma,i}$, one can stop the iteration as soon as the scaled isolation energy is exceeded.

For the purpose of this paper, the important point is that any prediction aiming to stay close to experimental fixed-cone photon definitions still requires a model or extraction of the photon fragmentation contribution \cite{Gehrmann:2022cih,Chen:2022gpk}.
This motivates the construction of photon fragmentation functions from \textsc{Pythia} and their representation in a portable gridded format.

\section{Improved interpolation of Photon Fragmentation Functions in \textsc{LHAPDF}}
\label{sec:lhapdf_interp}

\textsc{JetPhox}, \textsc{MCFM} and similar codes typically use two-dimensional linear interpolation between grid points to evaluate fragmentation functions at arbitrary values of $z$ and $\mu_D$.
Given the evolution \cref{eq:evoff}, the fragmentation functions can have non-trivial shapes, especially at low $z$ where the point-like contribution dominates.
\textsc{LHAPDF} 6.5.6 instead uses the default \texttt{logcubic} interpolator, \textit{i.e.} a bicubic Hermite spline in logarithmic coordinates of the independent variables.
For a generic tabulated fragmentation quantity $F_i(z,\mu_D^2)$, we define
\begin{equation}
	u = \log z
	\,,\qquad
	v = \log \mu_D^2
	\,.
\end{equation}
Inside one grid cell $(u_i,u_{i+1}) \times (v_j,v_{j+1})$, the interpolation is written as
\begin{equation}
	F_i(u,v)
	\simeq
	\sum_{m=0}^{3}\sum_{n=0}^{3} a^{(ij)}_{mn} \, t^m s^n
	\,,
	\qquad
	t = \frac{u-u_i}{u_{i+1}-u_i}
	\,,
	\qquad
	s = \frac{v-v_j}{v_{j+1}-v_j}
	\,.
\end{equation}
The sixteen coefficients $a^{(ij)}_{mn}$ are fixed by matching $F_i$, $\partial_u F_i$, $\partial_v F_i$, and $\partial_u \partial_v F_i$ at the four corners of the cell, as appropriate for bicubic Hermite interpolation.
Hence, \texttt{logcubic} does not mean taking the logarithm of the fragmentation function itself, but rather performing the interpolation in $\log z$ and $\log \mu_D^2$.
This is well suited to the approximately logarithmic scale dependence visible in \cref{eq:evoff} and to the steep small-$z$ behaviour.
If thresholds are present, \textsc{LHAPDF} can also use separate $\mu_D^2$ subgrids so that the interpolation does not cross discontinuities in the scale dependence \cite{Buckley:2014ana}.
We extract the photon fragmentation grid points of BFG~I and BFG~II from \textsc{JetPhox}, GRV~LO and GRV~NLO from \textsc{phoSec}~\cite{koenig:phd} as well as GDRG~LO and GDRG~NLO from \textsc{MCFM} and convert them into an \textsc{LHAPDF}-compatible format.
\Cref{tab:ffsets} summarizes the available sets and their properties.
\begin{table}[t]
\centering
\caption{
Compact summary of the photon fragmentation-function sets used in this work.
BFG~\cite{Bourhis:1997yu}  and GRV~\cite{Gluck:1992zx}  are VMD-based beyond leading logarithmic resummed parameterizations, while GDRG~\cite{Gehrmann-DeRidder:1997fom,Gehrmann-DeRidder:1998bju} denotes fixed-order quark-only fits to ALEPH data.
}
\label{tab:ffsets}
\small
\begin{tabular}{lcccc}
\toprule
Set & Accuracy & Scheme & $\mu_0^2$ [GeV$^2$] & Main feature \\
\midrule
BFG~I   & BLL/NLO & $\overline{\rm MS}$ & 2.0  & coherent VMD, smaller gluon FF \\
BFG~II  & BLL/NLO & $\overline{\rm MS}$ & 2.0  & coherent VMD, larger gluon FF \\
GRV~LO  & LL/LO   & $\mathrm{DIS}_\gamma$        & 0.3  & incoherent VMD, low-scale start \\
GRV~NLO & BLL/NLO & $\mathrm{DIS}_\gamma$        & 0.3  & incoherent VMD, low-scale start \\
GDRG~LO & LO fit  & $\overline{\rm MS}$                 & 0.14 & ALEPH-based quark-only fit \\
GDRG~NLO& NLO fit & $\overline{\rm MS}$                 & 0.64 & ALEPH-based quark-only fit \\
\bottomrule
\end{tabular}
\end{table}
Packaging the sets in \textsc{LHAPDF} is useful beyond convenience: it replaces process-specific interpolation code with a common interface, allows the photon-fragmentation input to be versioned and distributed like PDFs, and makes it immediately accessible to external tools such as \textsc{JetPhox}, \textsc{MCFM}, and \textsc{NNLOJet}.
Furthermore, it guarantees that the grids and interpolation are consistent across different codes, which is important for reproducibility and comparing results.
\textsc{LHAPDF} grids are also used to obtain the running coupling constant $\alpha_S$ at multiple loops precision.
Using this feature of \textsc{LHAPDF} with a photon FF is ill-advised as it is ill-defined.

\begin{figure}
        \centering
        \begin{subfigure}[t]{0.49\textwidth}
                \centering
                \includegraphics[width=\linewidth]{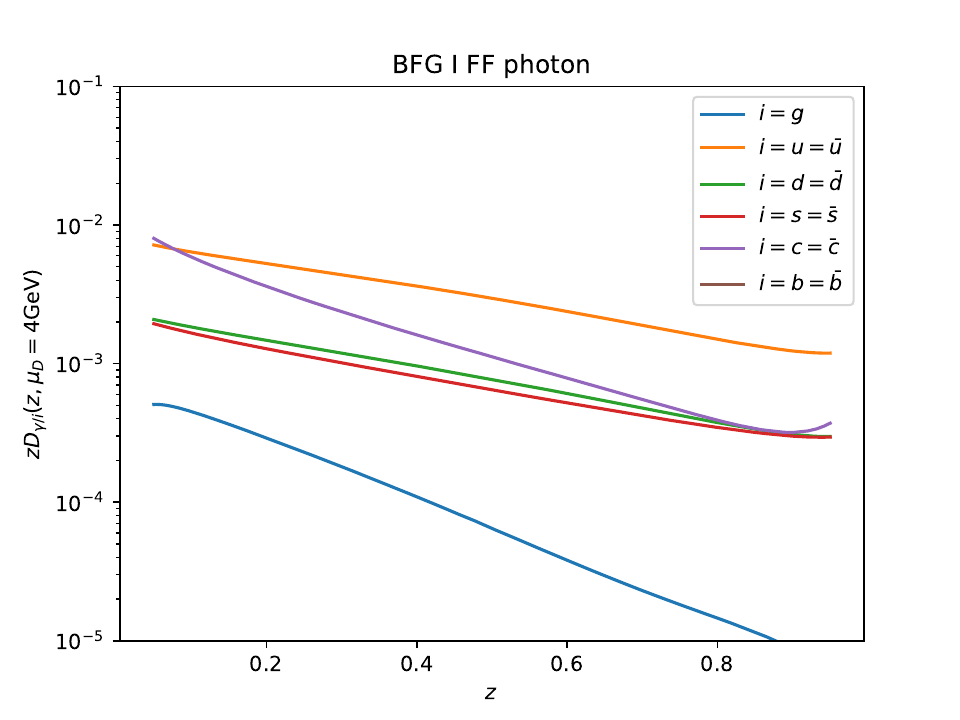}
                \caption{BFG~I.}
                \label{fig:bfg_i}
        \end{subfigure}
        \begin{subfigure}[t]{0.49\textwidth}
                \centering
                \includegraphics[width=\linewidth]{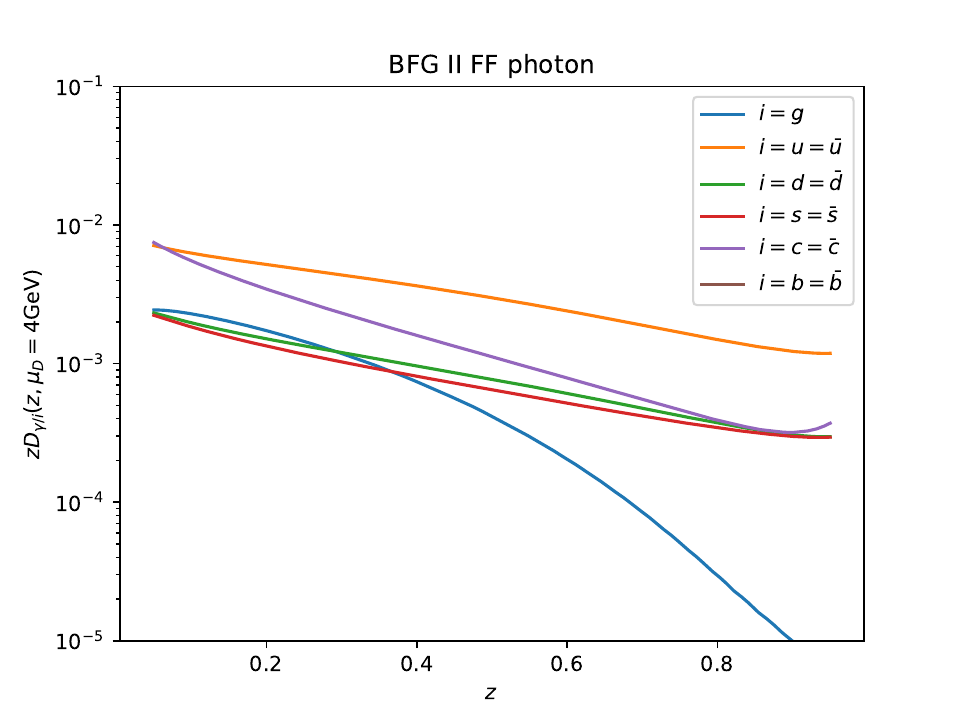}
                \caption{BFG~II.}
                \label{fig:bfg_ii}
        \end{subfigure}
        \begin{subfigure}[t]{0.49\textwidth}
                \centering
                \includegraphics[width=\linewidth]{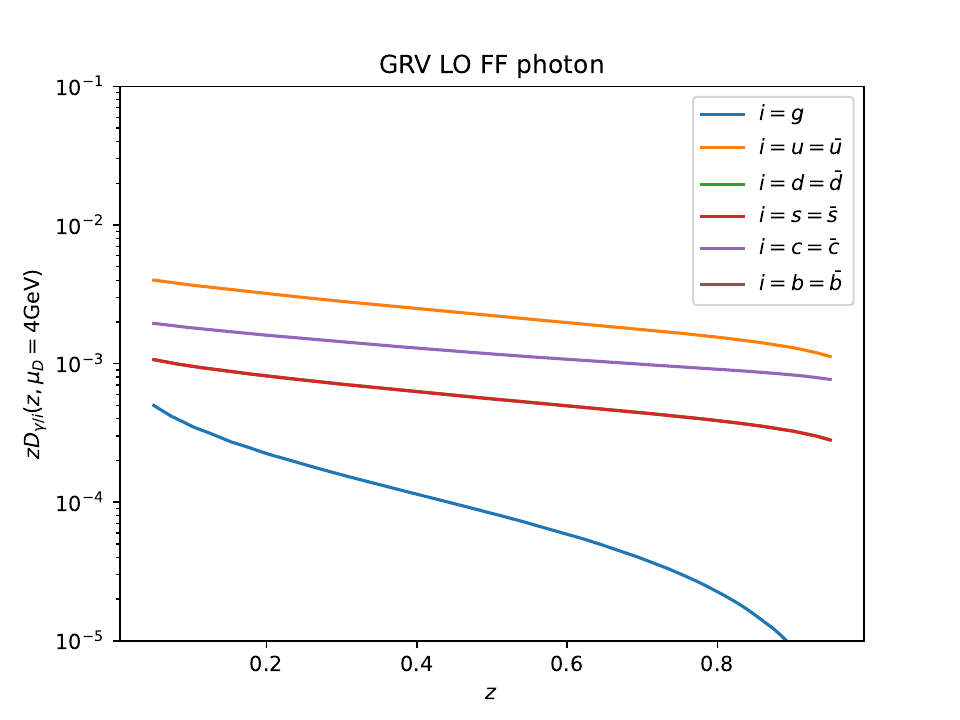}
                \caption{GRV~LO.}
                \label{fig:grv_lo}
        \end{subfigure}
        \begin{subfigure}[t]{0.49\textwidth}
                \centering
                \includegraphics[width=\linewidth]{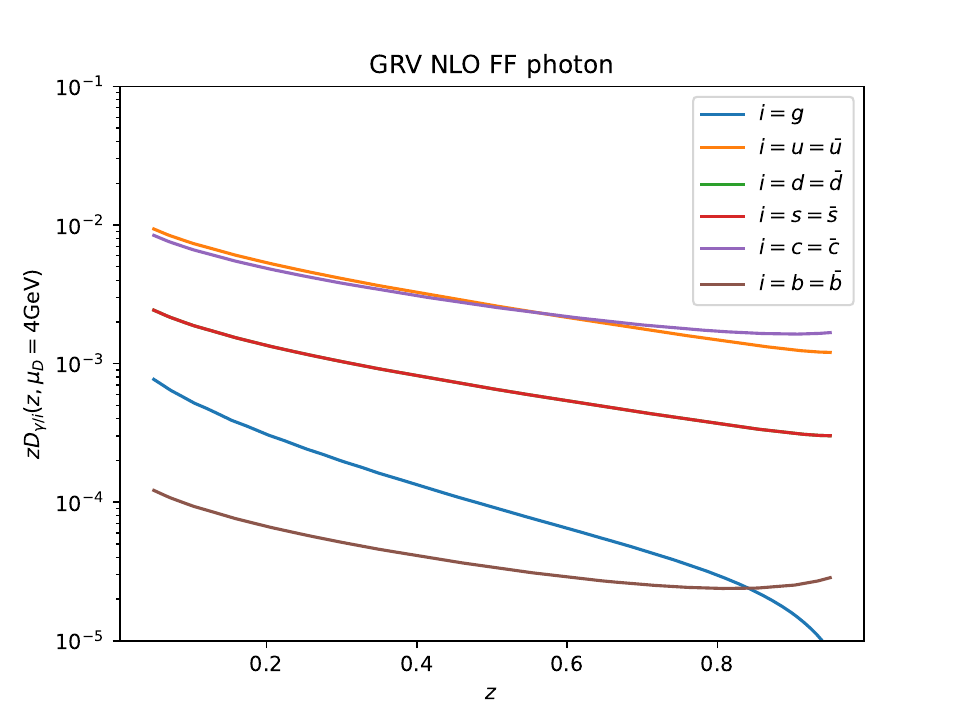}
                \caption{GRV~NLO.}
                \label{fig:grv_nlo}
        \end{subfigure}
        \begin{subfigure}[t]{0.49\textwidth}
                \centering
                \includegraphics[width=\linewidth]{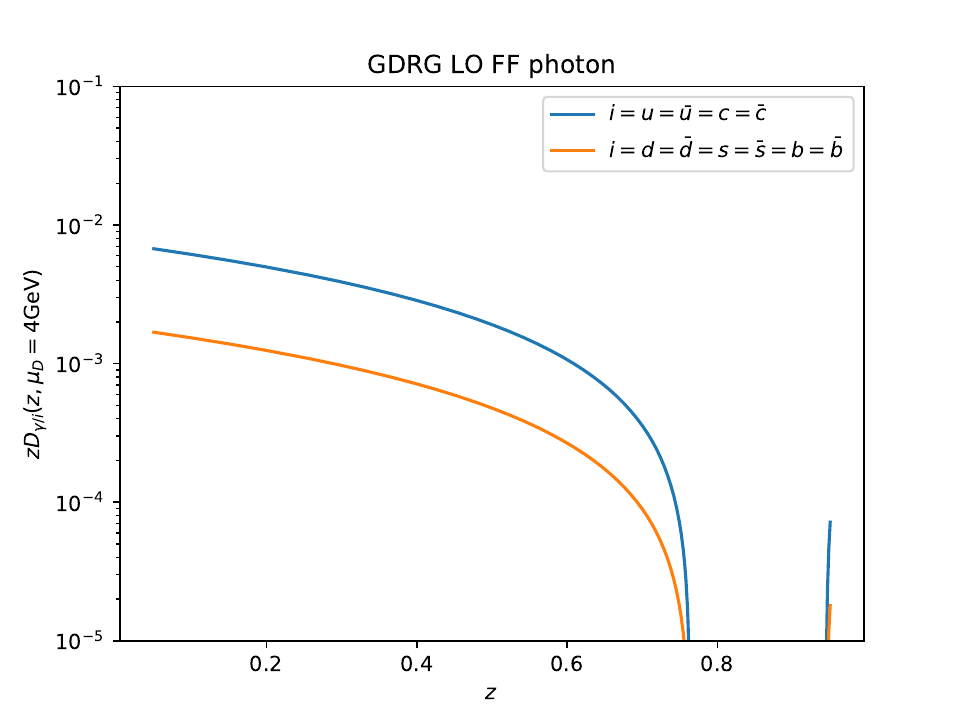}
                \caption{GDRG~LO.}
                \label{fig:gdrg_lo}
        \end{subfigure}\hfill
        \begin{subfigure}[t]{0.49\textwidth}
                \centering
                \includegraphics[width=\linewidth]{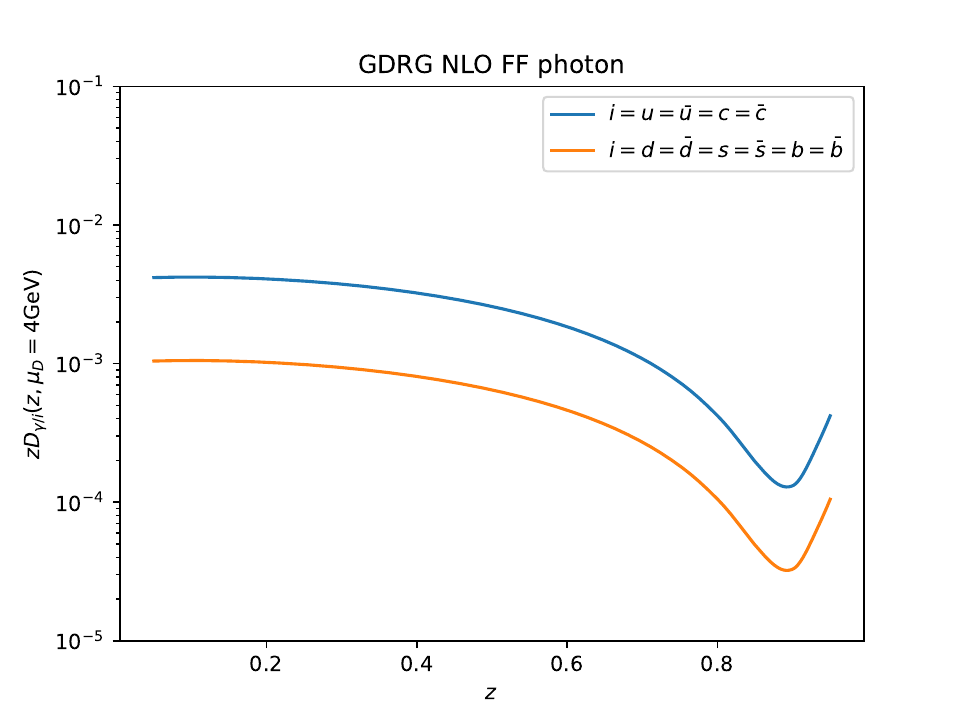}
                \caption{GDRG~NLO.}
                \label{fig:gdrg_nlo}
        \end{subfigure}
        \caption{Photon fragmentation functions at $\mu_D = 4\,\mathrm{GeV}$ for the BFG~I, BFG~II, GRV~LO, GRV~NLO, GDRG~LO, and GDRG~NLO parameterizations.}
        \label{fig:ff_models}
\end{figure}

\Cref{fig:ff_models} shows BFG~I, BFG~II, GRV~LO, GRV~NLO, GDRG~LO, and GDRG~NLO fragmentation functions at $\mu_D = 4\,\mathrm{GeV}$ as a function of $z$.
They do not differentiate between quark and anti-quark flavours.
The different sets exhibit sizeable variations both in their overall normalization and in their flavour dependence, especially for the charm and bottom flavours whose mass is close to the fragmentation scale.
Furthermore, GDRG does not include a gluon component, since it was developed for $e^+ e^-$ collisions, where the contribution from gluon fragmentation was assumed to be negligible.
The most surprising curve here is GDRG~LO which turns to very low values for all quarks at $z\approx 0.9$.
It needs to be emphasized here that photon FFs are factorization-scheme-dependent quantities, not directly observable probabilities and thus can become negative in certain regions of phase space.
In combination with the corresponding partonic cross sections, the physical cross section is always positive and well-defined.

\begin{figure}
        \centering
        \begin{subfigure}[t]{0.49\textwidth}
        \centering
                \includegraphics[width=\textwidth]{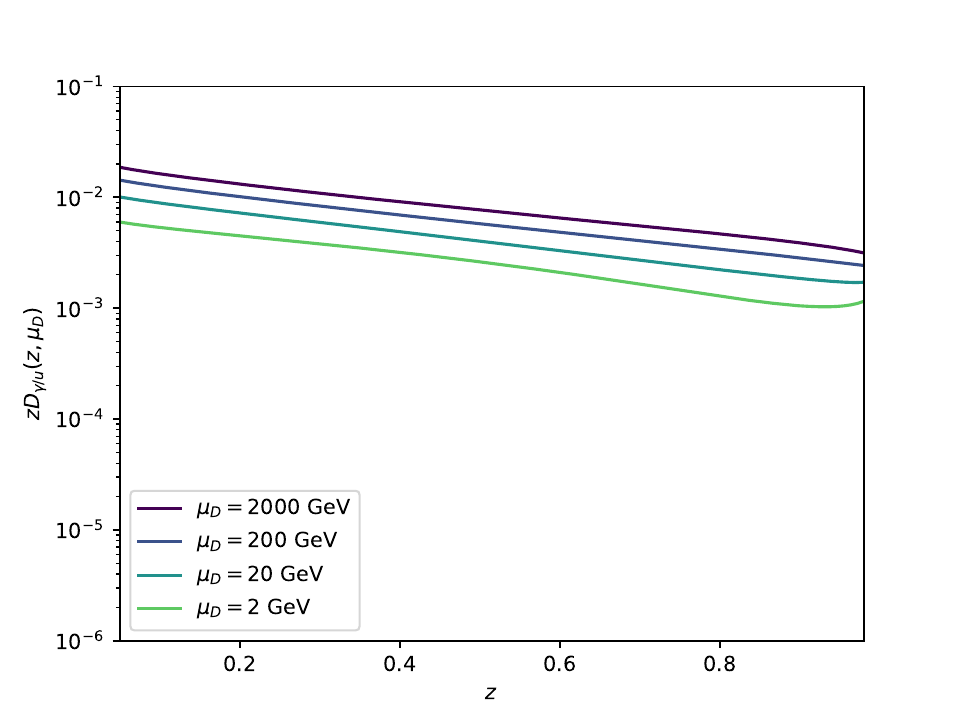}
                \caption{
                        Up quark.
                }
        \end{subfigure}
        \begin{subfigure}[t]{0.49\textwidth}
        \centering
                \includegraphics[width=\textwidth]{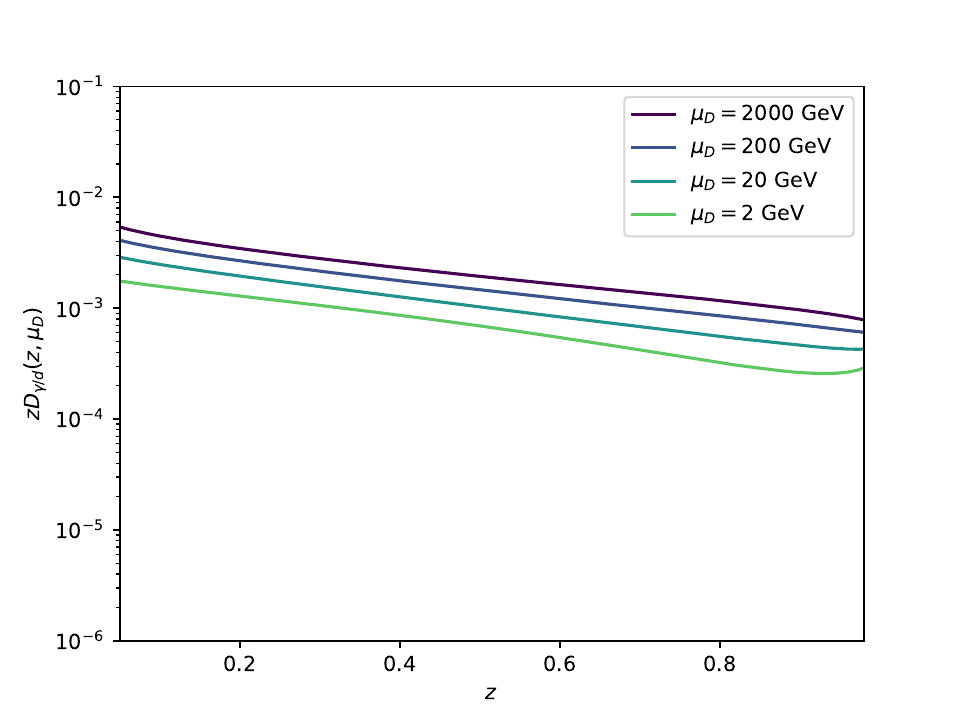}
                \caption{
                        Down quark.
                }
        \end{subfigure}
        \begin{subfigure}[t]{0.49\textwidth}
        \centering
                \includegraphics[width=\textwidth]{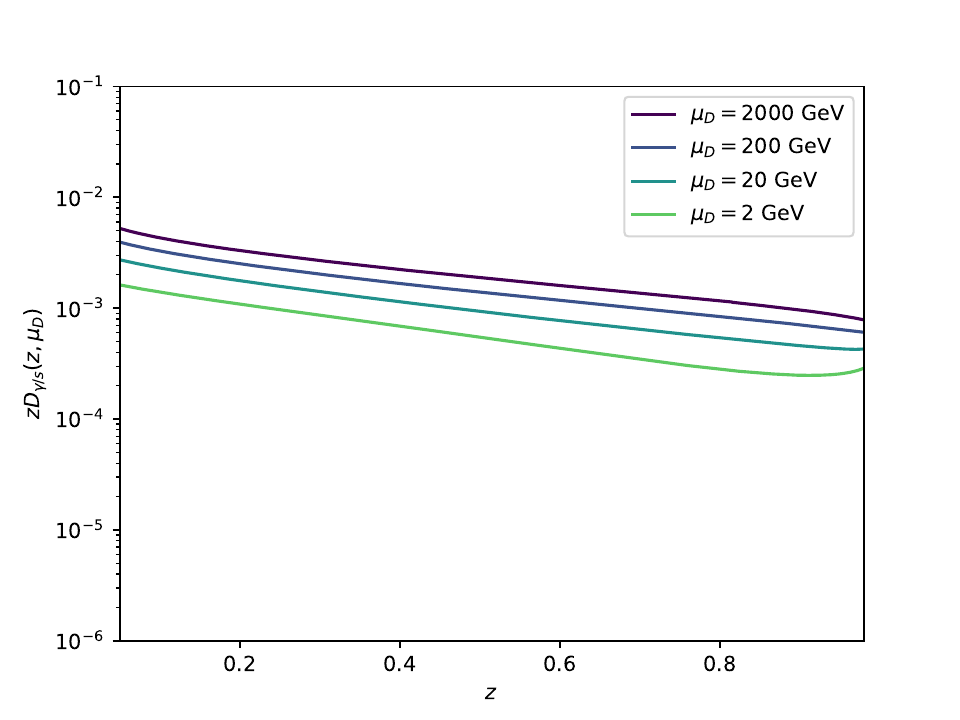}
                \caption{
                        Strange quark.
                }
        \end{subfigure}
        \begin{subfigure}[t]{0.49\textwidth}
        \centering
                \includegraphics[width=\textwidth]{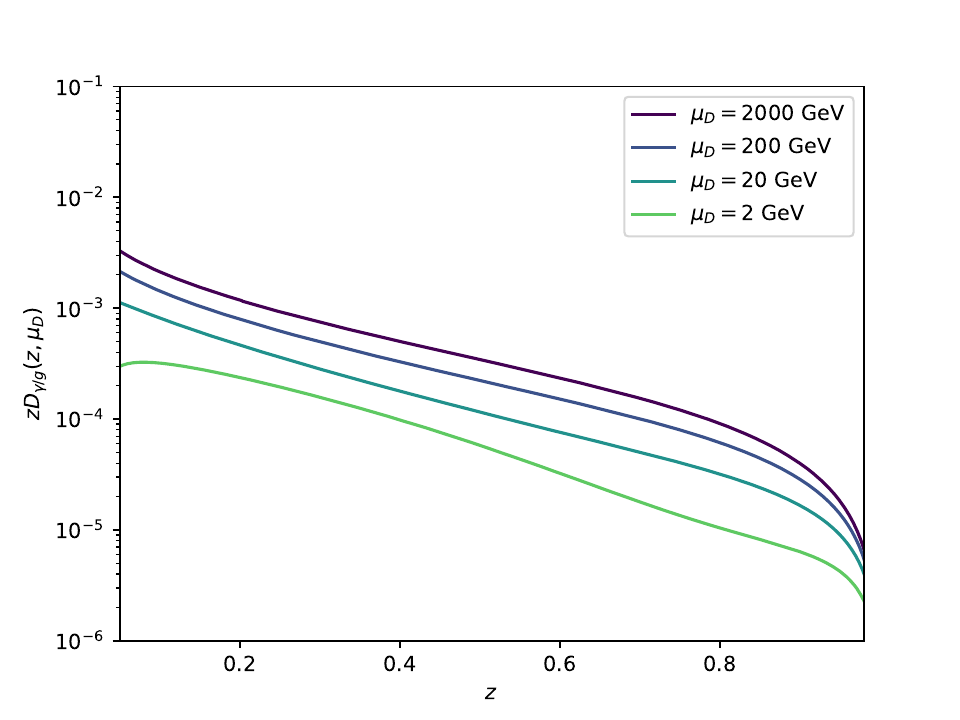}
                \caption{
                        Gluon.
                }
        \end{subfigure}
        \caption{
                BFG~I FF extracted from \textsc{JetPhox} and linear interpolation by \textsc{LHAPDF} of the up quark, down quark, strange quark and gluon as a function of $z$ and $\mu_D$.
        }
        \label{fig:all_bfg_i}
\end{figure}

To show the effect of evolution in the fragmentation scale $\mu_D$, \cref{fig:all_bfg_i} shows the BFG~I fragmentation functions as a function of $z$ for different fragmentation scales $\mu_D$ for the up quark, down quark, strange quark and gluon.
The largest contribution comes from the up quark while the down and strange quark contributions are roughly the same.
The gluon contribution is the lowest, as expected, since it only plays a significant role beyond LO.
As the scale increases, there is a larger contribution from the parton fragmenting into a photon as expected from \cref{eq:evoff}.
This can already be connected to the parton-shower picture, where the scale is evolved downwards and the parton can radiate a photon at any stage of the shower.
Hence, a larger scale allows for more radiation and therefore a higher yield of photon fragmentation.

In order to achieve compatibility with existing codes, we provide both the legacy \texttt{linear} interpolation grids (e.g. \texttt{BFG\_I}) and improved \texttt{logcubic} grids (e.g. \texttt{BFG\_I\_improved}) in \textsc{LHAPDF} (see \cref{tab:sets}) of the BFG and GRV families.
Further, the legacy linear interpolations have hard cuts down to 0 at the end of the tabulated $z$ range.
This behaviour is reproduced in the legacy linear interpolations by manually adding points quickly approaching 0.
By contrast, the \texttt{logcubic} mode allows for a smooth extrapolation outside the tabulated $z$ range, with boundaries going to 0 for $z \to 0$ and $z \to 1$, most prominently both BFG~I and II are $D_{\gamma/i}(z \lesssim 0.05, \mu_D)=0$.
For GDRG, which does not rely on a grid or interpolation in \textsc{MCFM}, instead a sufficiently dense grid was evaluated to construct a robust \texttt{logcubic} interpolation, which is then provided in \textsc{LHAPDF}.

\begin{figure}
        \centering
        \begin{subfigure}[t]{0.49\textwidth}
        \centering
                \includegraphics[width=\textwidth]{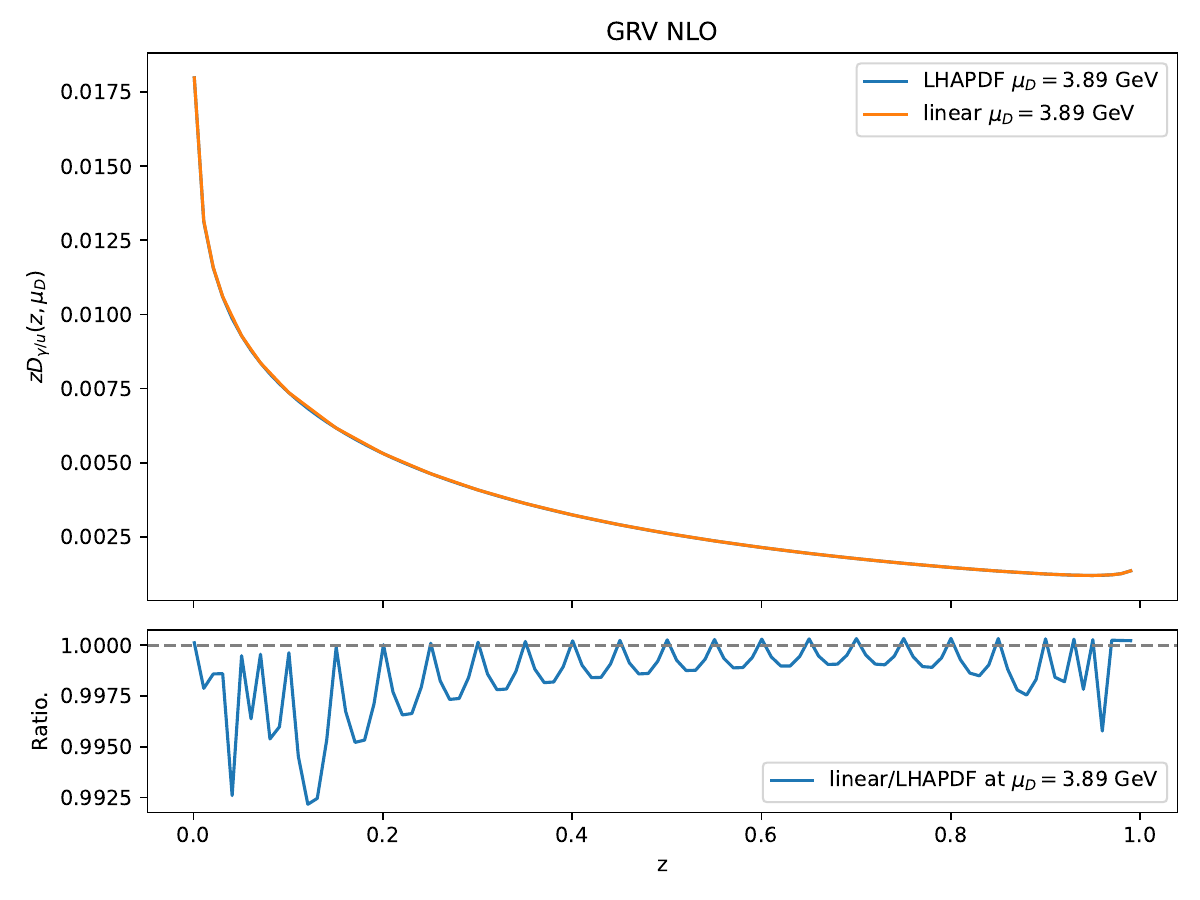}
                \caption{
                        Up quark.
                }
        		\label{fig:comparison_Q:grv_nlo_upquark}
        \end{subfigure}
        \begin{subfigure}[t]{0.49\textwidth}
        \centering
                \includegraphics[width=\textwidth]{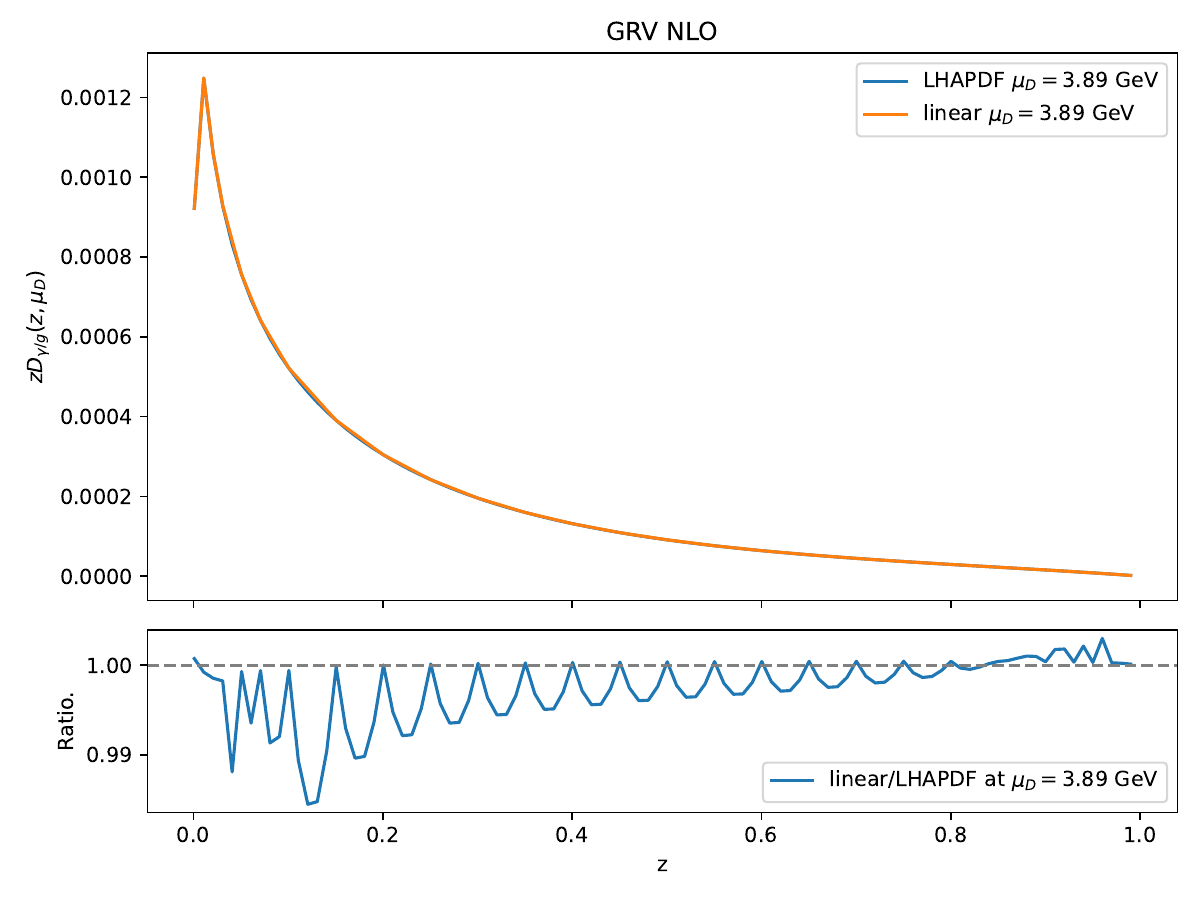}
                \caption{
                        Gluon.
                }
        		\label{fig:comparison_Q:grv_nlo_gluon}
        \end{subfigure}
                \begin{subfigure}[t]{0.49\textwidth}
        \centering
                \includegraphics[width=\textwidth]{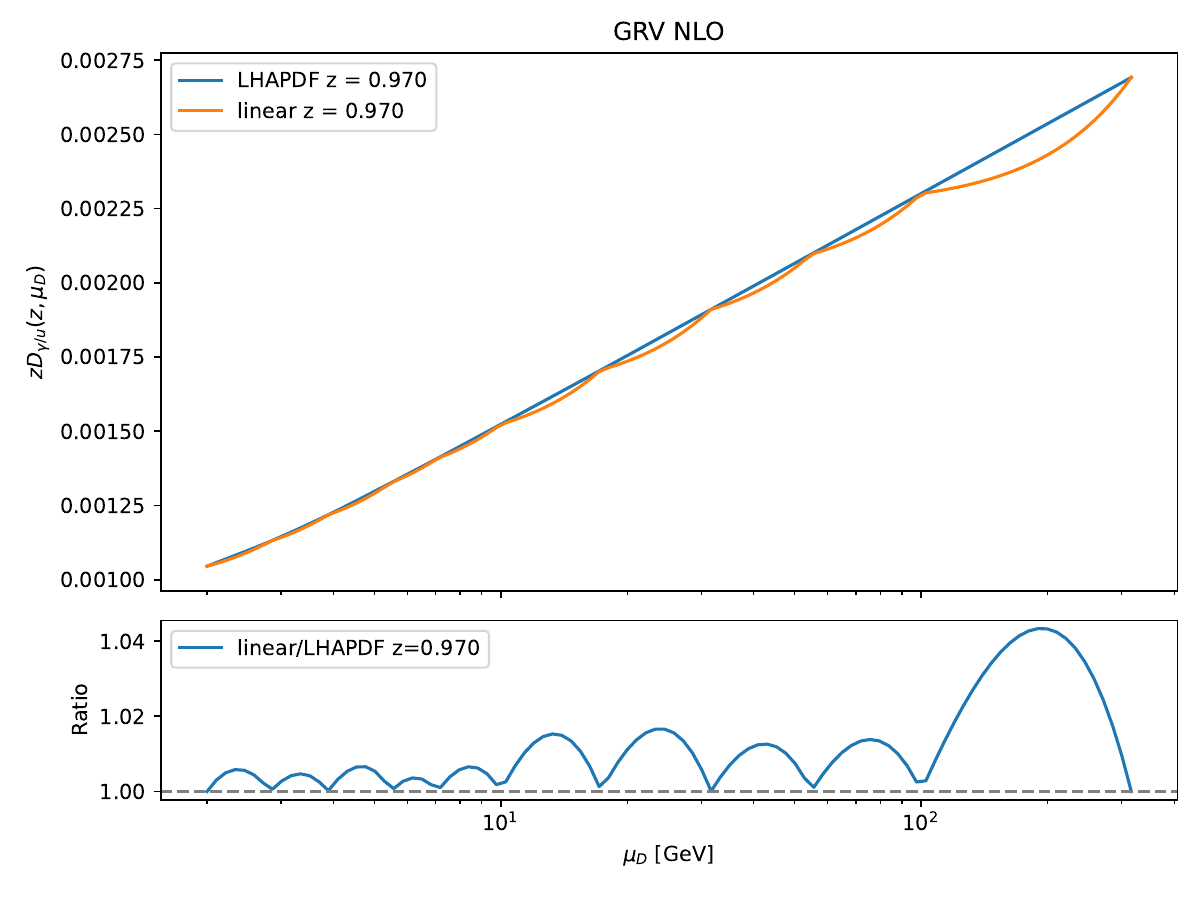}
                \caption{
                        Up quark.
                }
        		\label{fig:comparison_z:grv_nlo_upquark}
        \end{subfigure}
        \begin{subfigure}[t]{0.49\textwidth}
        \centering
                \includegraphics[width=\textwidth]{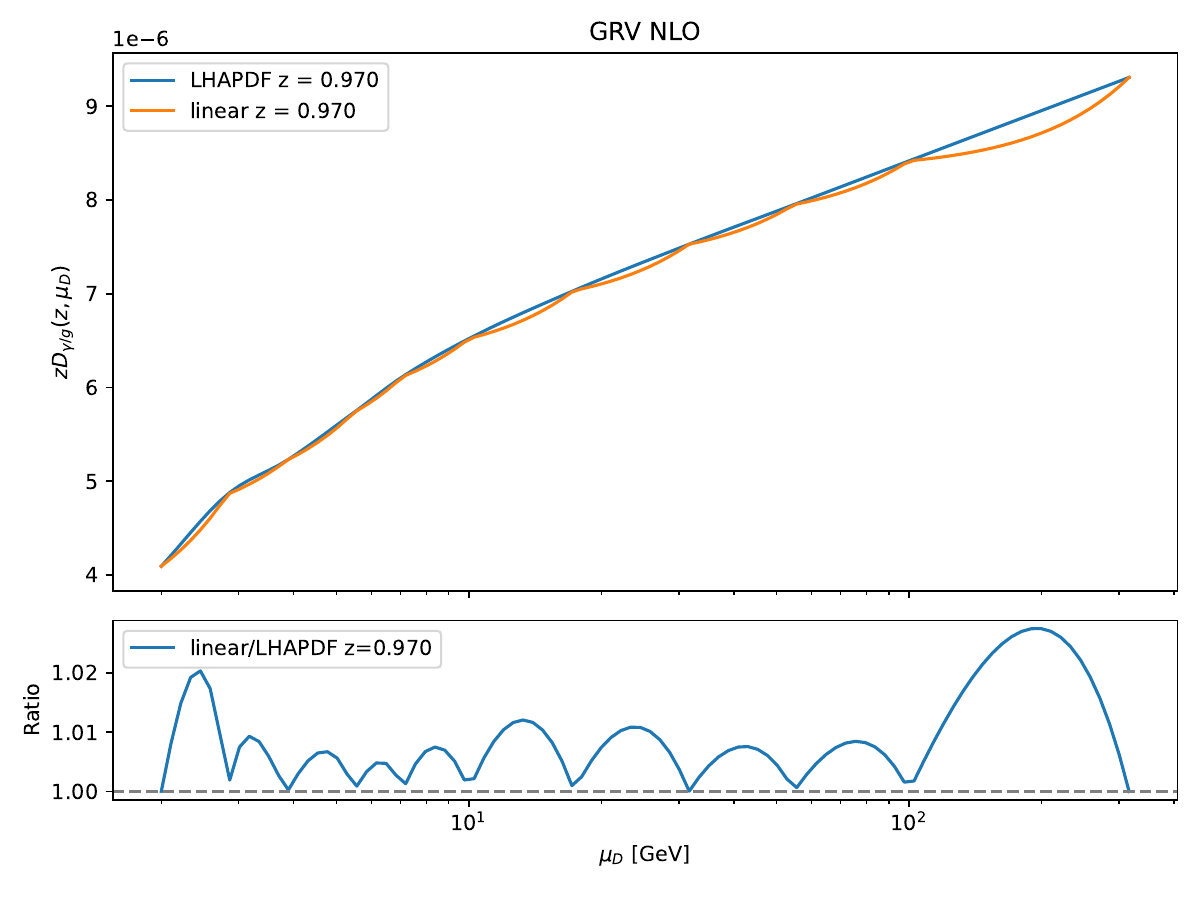}
                \caption{
                        Gluon.
                }
        		\label{fig:comparison_z:grv_nlo_gluon}
        \end{subfigure}
        \caption{
                GRV~NLO FF and interpolation by \textsc{LHAPDF} of the up quark and gluon as a function of $z$ and $\mu_D$.
        }
        \label{fig:comparison}
\end{figure}

In \cref{fig:comparison} we compare the linear interpolation with the \texttt{logcubic} interpolation of \textsc{LHAPDF} for the GRV~NLO fragmentation functions.
In both \cref{fig:comparison_Q:grv_nlo_upquark} and \cref{fig:comparison_Q:grv_nlo_gluon} we see a repeated increasing and then shrinking deviation between the two interpolation methods as a function of $z$ for the up quark and gluon.
The points of the least deviation correspond to the grid points where the linear interpolation is exact.
The discrepancy between the two interpolation methods is always less than 1\% and therefore negligible for phenomenological applications.

In \cref{fig:comparison_z:grv_nlo_upquark} and \cref{fig:comparison_z:grv_nlo_gluon} instead we show the comparison as a function of $\mu_D$ while keeping $z$ fixed.
We see that the linear interpolation does not capture the expected logarithmic scale dependence of the fragmentation functions, while the \texttt{logcubic} interpolation of \textsc{LHAPDF} does a much better job at reproducing the expected smooth behaviour.
All observations above also apply to GRV~LO, BFG~I and II.
However, the deviation behaviour for BFG is less dramatic, since it uses more grid points in $\mu_D$ and hence the linear interpolation is less coarse.

\begin{figure}
        \centering
        \begin{subfigure}[t]{0.49\textwidth}
        \centering
                \includegraphics[width=\textwidth]{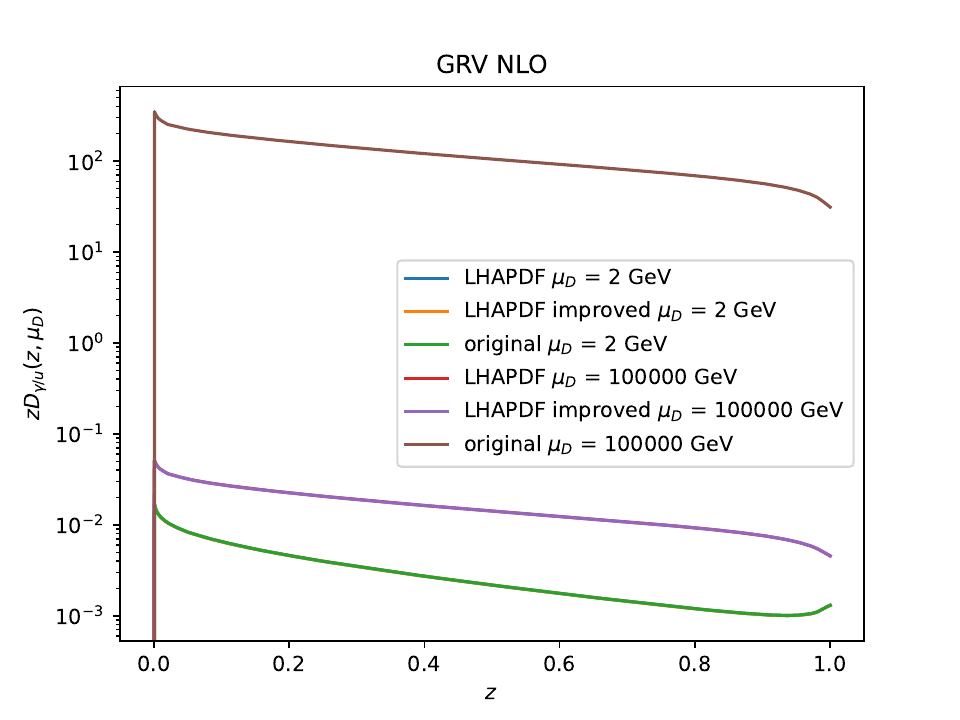}
                \caption{
                        GRV~NLO
                }
                \label{fig:extrapolation:grv_nlo_upquark}
        \end{subfigure}
        \begin{subfigure}[t]{0.49\textwidth}
        \centering
                \includegraphics[width=\textwidth]{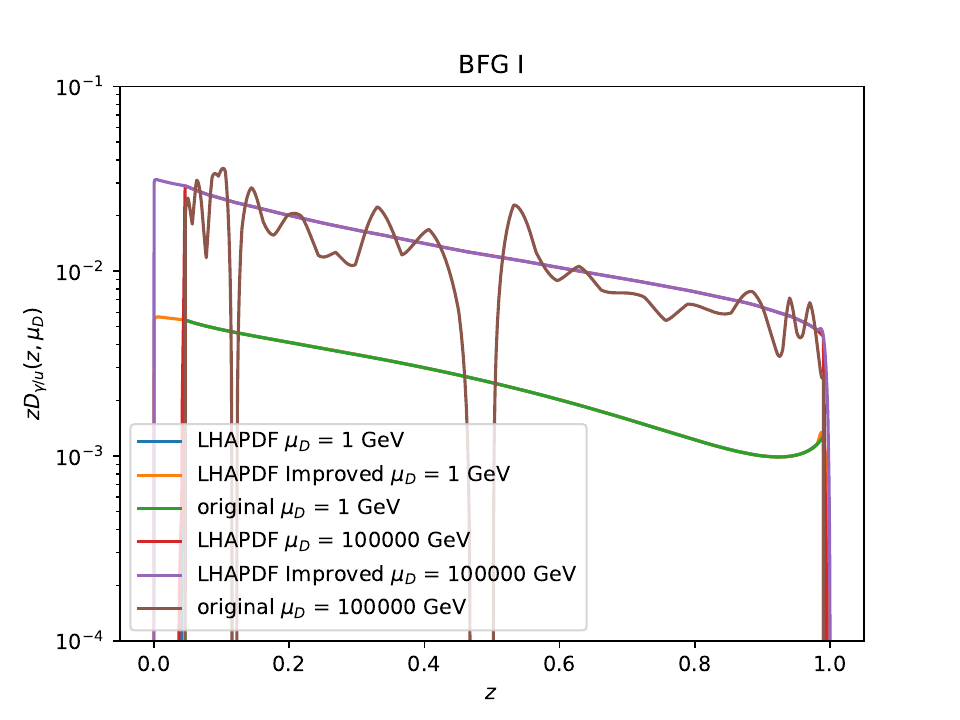}
                \caption{
                        BFG~I
                }
                \label{fig:extrapolation:bfg_i_upquark}
        \end{subfigure}
        \caption{
                BFG~I and GRV~NLO FF extrapolation in $\mu_D$ between the original linear and new LHAPDF codes for the up quark.
        }
        \label{fig:extrapolation}
\end{figure}

\Cref{fig:extrapolation} shows the extrapolation in $\mu_D$ of the BFG~I and GRV~NLO FFs between the original linear and new \textsc{LHAPDF} codes.
The first observation is that in \cref{fig:extrapolation:grv_nlo_upquark} the value of $zD(z,\mu_D=100$ TeV$)$ exceeds 1 by multiple orders, signalling an unstable extrapolation.
Both linear and logcubic of \textsc{LHAPDF} extrapolations instead give reasonable values for $zD(z,\mu_D=100$ TeV$)$.
\Cref{fig:extrapolation:bfg_i_upquark} does not show the same unphysical behaviour, but suffers from numerical instabilities at very large values of $\mu_D=100$ TeV.
While the figures only show the up quark, the same behaviour is observed for all other quark flavours and the gluon.
Furthermore, in the case of the BFG~I improved FF we see that the interpolation of \textsc{LHAPDF} beyond the original cut-off gives reasonable values for $zD(z \lesssim 0.04,\mu_D) \neq 0$.

\section{Extraction of Photon Fragmentation Functions from \textsc{Pythia}}

Having established a common \textsc{LHAPDF} representation of the conventional photon-fragmentation parameterizations and clarified the role of interpolation, we now turn to a shower-based extraction of photon fragmentation functions from \textsc{Pythia} 8.317.
The aim is not to replace the existing analytic parameterizations, but to construct a practical baseline in the same $(z,\mu_D)$ variables so that the flavour structure and scale dependence can be compared directly.
We first define the extraction procedure, then describe how samples from different hard-process transverse-momentum slices are combined to populate the full kinematic plane, and finally compare the resulting distributions with the conventional fragmentation-function sets.

\subsection{Methodology}
\label{sec:methodology}

Our objective is to construct practical shower-level effective photon fragmentation functions from \textsc{Pythia} and to represent the result on a grid that can be consumed in the same way as the analytic sets.
To distinguish it from the renormalized, scheme-dependent fragmentation functions $D_{\gamma/i}(z,\mu_D)$ entering fixed-order factorization, we denote the extracted shower quantity by $D^{\mathrm{PS}}_{\gamma/i}(z,\mu_D)$.
We nevertheless tabulate it in the same $(z,\mu_D)$ variables to enable pragmatic phenomenological comparisons with the conventional sets.
For each chosen fragmentation scale $\mu_D$, we analyze the final-state photons generated by the shower and build binned distributions in the photon momentum fraction $z$ for each initiating parton flavour.
Thus,
\begin{align}
        D^{\mathrm{PS}}_{\gamma/i}(z,\mu_D) &= \frac{1}{\sigma_{i \to Y}(\mu_D) + \sigma_{i \to \gamma + X}(\mu_D)} \frac{\dd \sigma_{i \to \gamma + X}(z,\mu_D)}{\dd z} 
        \\&= \frac{1}{N_{i,\text{tot}}(\mu_D)} \frac{\dd N_{i \to \gamma + X}(z,\mu_D)}{\dd z}
\end{align}
where $\mu_D$ is the fragmentation scale identified as $p_{T,\text{parton}} = p_{T,i}$ of the parton $i$ in the hard process, $X$ can be anything and $Y$ does not contain photons\footnote{The reason behind picking this scale instead of the more conventional $p_{T,\gamma}$ is that the no-photon case does not provide a matching fragmentation scale then.}.
The longitudinal momentum fraction $z$ is determined via
\begin{equation}
        z = z_\text{light-cone} = \frac{ E_\gamma + \vec{p}_\gamma \cdot \hat{n}_i }{ E_i + \vec{p}_i \cdot \hat{n}_i }
\end{equation}
with $\hat{n}_i = \frac{\vec{p}_i}{|\vec{p}_i|}$, where $\vec p_i$ is the momentum of the initiating parton and $\vec p_\gamma$ is the momentum of the photon.
Alternative definitions of $z$ are possible, i.e. $z_\text{long} =\frac{\vec p_i \cdot \vec p_\gamma}{ \vec p_i \cdot \vec p_i }$.
While the longitudinal and light-cone definitions of the momentum fraction differ for finite-angle emissions, they coincide in the collinear limit relevant to the definition of the fragmentation function.

In practical terms, we use \textsc{Pythia}'s \verb|HardQCD:qqbar2gg|, \verb|HardQCD:gg2bbbar|, \verb|HardQCD:gg2ccbar| or \verb|HardQCD:gg2qqbar| processes with an \verb|UserHook| to restrict the final state to a given final state flavour.
Only final state radiation is permitted and the photon emission is enhanced by a factor of 50 and absorbed in a reduced event weight, since otherwise the statistics are too low.
Within the \textsc{HepMC}~\cite{Dobbs:2001ck,Verbytskyi:2020sus} event record the photon can be traced back to the initiating parton, which allows us to compute the fragmentation functions.
Neither Multi-Parton-Interactions nor hadronization nor decays are included, so that the extraction is performed at the parton-shower level.
In the approach pursued here, no explicit VMD-type non-perturbative input is introduced.
Consequently, the extracted FF should be interpreted as the perturbatively generated contribution to photon fragmentation, with the low-scale boundary condition $D^\text{np}_{\gamma/i}(z,\mu_0)$ in \cref{eq:evoff} effectively replaced by the shower-cutoff prescription of \textsc{Pythia}.

\begin{figure}
        \centering
        \begin{subfigure}[t]{0.49\textwidth}
                \centering
                \includegraphics[width=\linewidth]{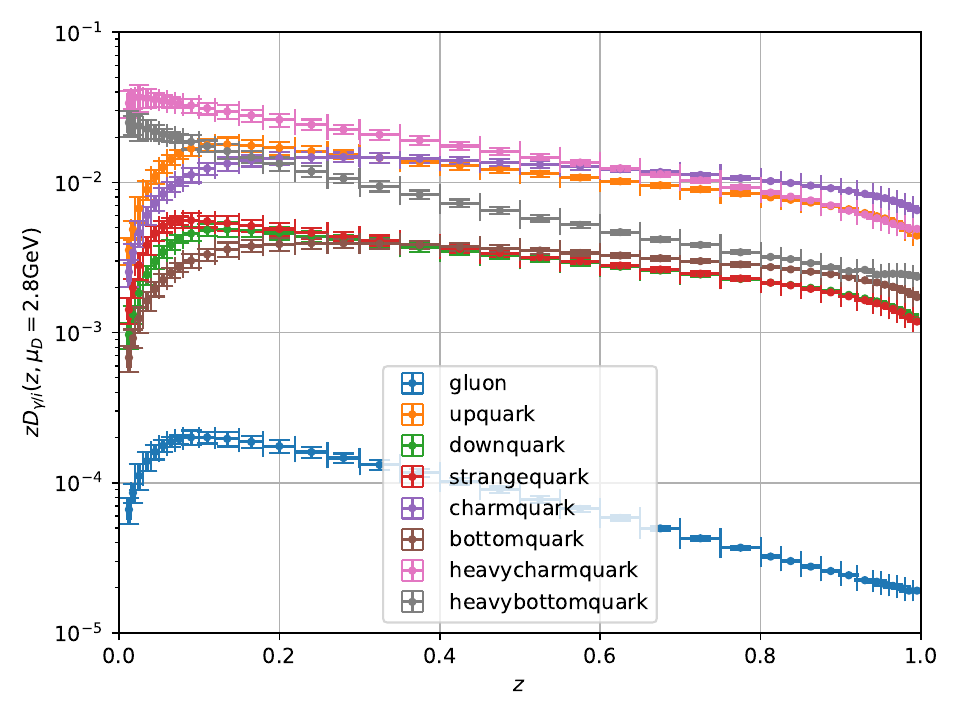}
                \caption{$\mu_D = 2.8\,\mathrm{GeV}$.}
        \end{subfigure}
        \begin{subfigure}[t]{0.49\textwidth}
                \centering
                \includegraphics[width=\linewidth]{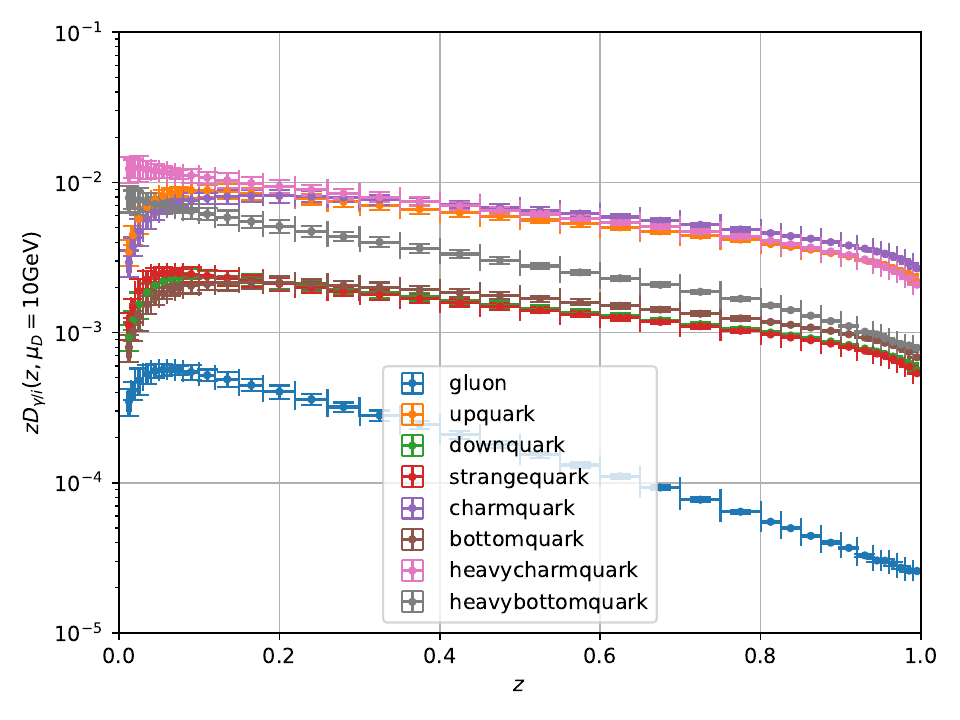}
                \caption{$\mu_D = 10\,\mathrm{GeV}$.}
        \end{subfigure}
        \begin{subfigure}[t]{0.49\textwidth}
                \centering
                \includegraphics[width=\linewidth]{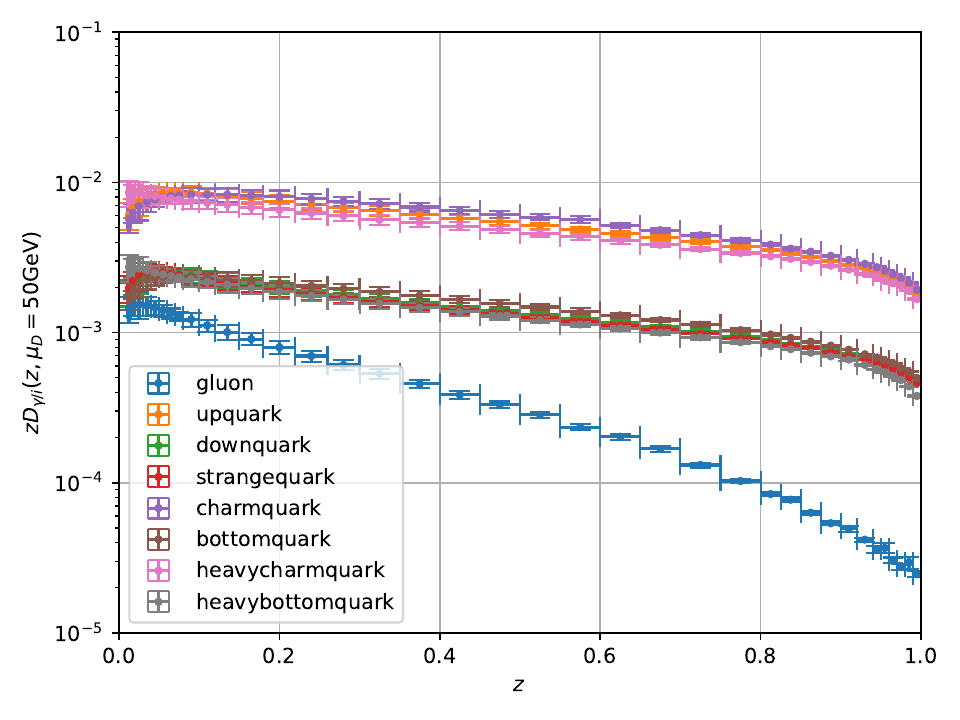}
                \caption{$\mu_D = 50\,\mathrm{GeV}$.}
        \end{subfigure}
        \begin{subfigure}[t]{0.49\textwidth}
                \centering
                \includegraphics[width=\linewidth]{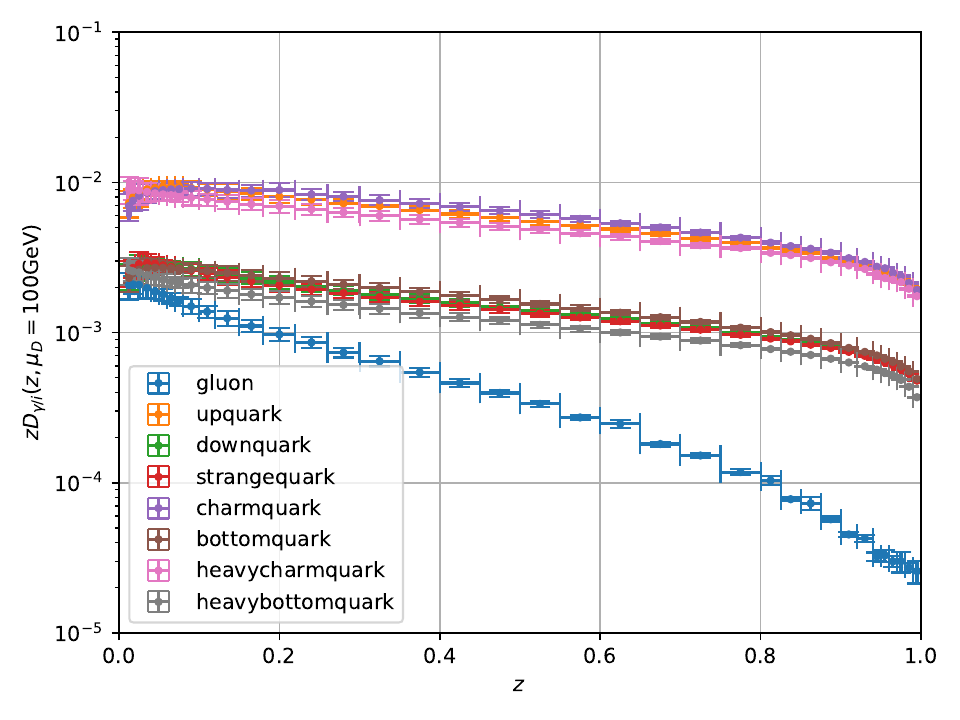}
                \caption{$\mu_D = 100\,\mathrm{GeV}$.}
        \end{subfigure}
        \caption{\textsc{Pythia}-extracted photon fragmentation distributions for the gluon and quark channels at representative fragmentation scales as a function of $z$.}
        \label{fig:pythia_ff}
\end{figure}

The resulting histograms provide a Monte-Carlo estimator for $D^{\mathrm{PS}}_{\gamma/i}(z,\mu_D)$, including the gluon and light- and heavy-quark channels shown below.
For readability, we omit the superscript $\mathrm{PS}$ in the following.
The extracted flavour channels are then tabulated in the same variables $(z,\mu_D)$ used by conventional photon-fragmentation parameterizations, which makes a direct comparison and an \textsc{LHAPDF} export straightforward.
Representative extracted distributions are shown in \cref{fig:pythia_ff}.

The quark channels dominate over the gluon channel, with the up-quark contribution largest because of its larger electric charge.
As $\mu_D$ increases, the massless fragmentation yield grows, in qualitative agreement with the leading-order expectation in \cref{eq:evoff} and \cref{fig:all_bfg_i}.
The massive bottom and charm quark instead decrease as $\mu_D$ increases.
At large scales the difference between the massless and massive quark channels is small.
In particular, the shower reproduces the expected hierarchy in which the gluon contribution remains comparatively small and is populated only indirectly through QCD mixing.
While we show also results for simulating the hard process in \textsc{Pythia} for massive charm and bottom quarks in the following we will treat all quarks as massless.
This is different to the treatment in Ref.~\cite{Klasen:2014xfa} where the massive modes only appear when the scale $\mu_D$ is above the mass threshold.
The dominant difference between the massive and the massless case is an increase in photon emissions at low $z$ and low $\mu_D$.
In the region $\mu_D \sim m_Q$, the heavy-flavour channels are less reliable because threshold effects and potentially large logarithms of $\mu_D/m_Q$ would require a more dedicated treatment than is intended here.
For the present exploratory study, the massless setup is therefore sufficient and has the additional advantage of being more directly comparable to the existing photon-fragmentation parameterizations.

Since it is difficult to populate the full $(z,\mu_D)$ plane with a single \textsc{Pythia} run, we combine multiple runs with different hard-process transverse-momentum slices (see \cref{sec:ptslices}).
While in principle only one distribution at fixed $\mu_D$ needs to be extracted and then evolved to other scales, to the authors' best knowledge there is no public implementation of the evolution equations for photon fragmentation functions, so we extract the distributions at multiple scales.

\subsection{Parton shower parameter dependency}

The shower-extracted fragmentation functions should be interpreted as model-dependent inputs, so it is important to identify which aspects are robust and which remain tune or algorithm dependent.
The dominant structure is controlled by the flavour of the initiating parton, the momentum fraction $z$, and the fragmentation scale $\mu_D$.
Because the extraction is formulated in terms of local shower branchings, any residual dependence on the hadronic center-of-mass energy is expected to be subleading once the relevant phase space is populated.
We performed the extraction at both 5 TeV and 13 TeV and found no significant differences in the resulting distributions, confirming that the extraction is indeed largely independent of the hard-process energy.
Therefore, all results shown in the following are obtained at 13 TeV.

Other systematic variations are also worth exploring.
In particular, it would be valuable to compare the default \textsc{Pythia} shower with alternative shower formalisms such as \textsc{Vincia}, and to study how much the extracted distributions change when one moves from the partonic definition adopted here to a jet-based definition closer to photon-in-jet observables.
We therefore regard the present extraction as a baseline shower result on top of which such systematic studies can be added.

At small $z$ the shower-extracted fragmentation functions exhibit a suppression relative to conventional parameterizations.
This behaviour is expected to be sensitive to the infrared cut-off of the parton shower.
In \textsc{Pythia}, QED radiation from quarks is terminated at a finite transverse-momentum scale, controlled by \verb|TimeShower:pTminChgQ|, which restricts the available splitting phase space and becomes increasingly relevant at small $z$ and low fragmentation scales.
By reducing it to the allowed minimum we find that the low-$z$ suppression is reduced, but not eliminated.


\subsection{Comparison between \textsc{Pythia} and Photon Fragmentation Functions}
\label{sec:comparison}

\begin{figure}
        \centering
        \begin{subfigure}[t]{0.49\textwidth}
        \centering
                \includegraphics[width=\textwidth]{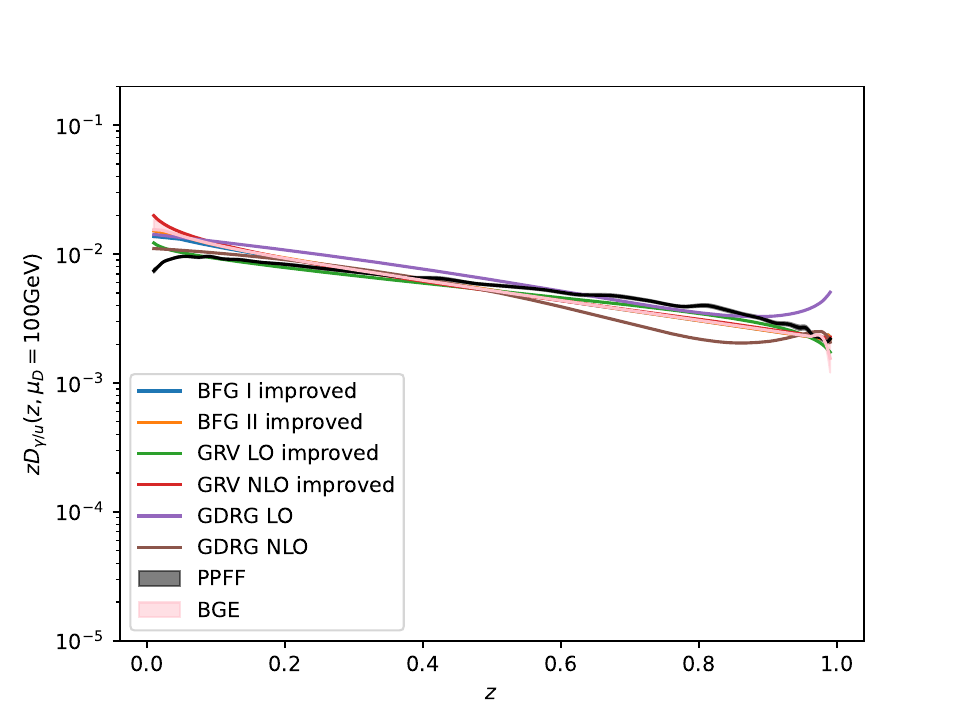}
                \caption{
                        Up quark.
                }
        \end{subfigure}
        \begin{subfigure}[t]{0.49\textwidth}
        \centering
                \includegraphics[width=\textwidth]{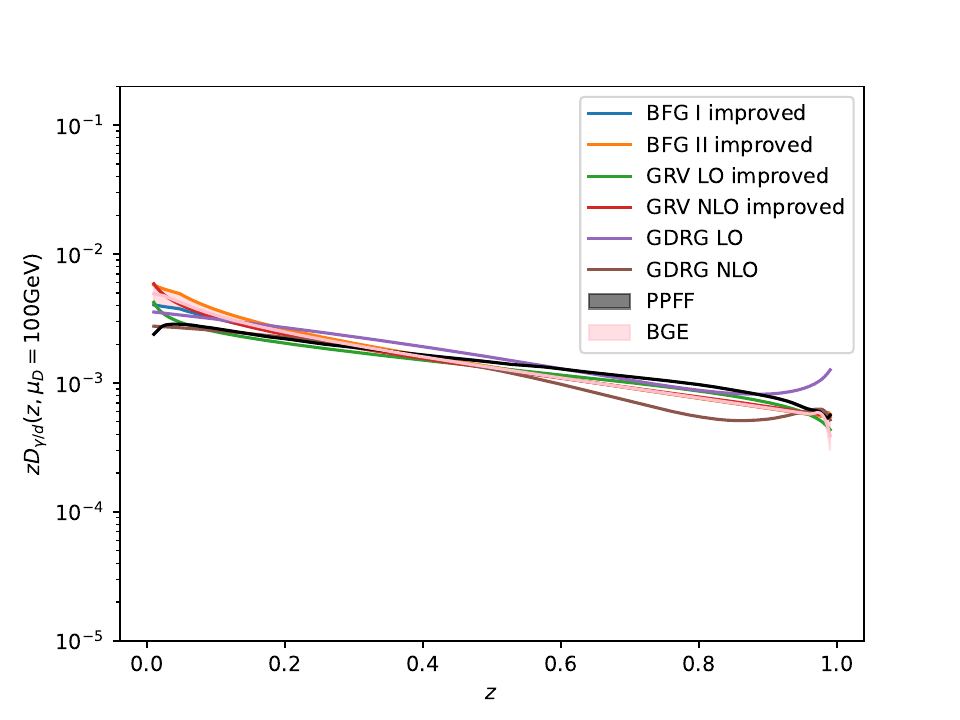}
                \caption{
                        Down quark.
                }
        \end{subfigure}
        \begin{subfigure}[t]{0.49\textwidth}
        \centering
                \includegraphics[width=\textwidth]{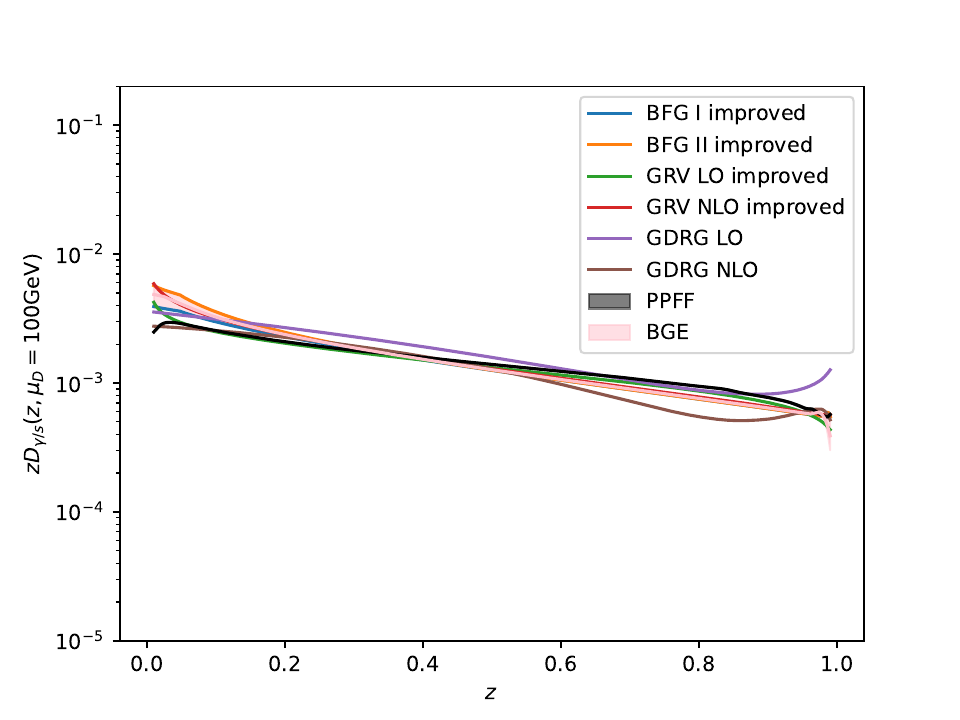}
                \caption{
                        Strange quark.
                }
        \end{subfigure}
        \begin{subfigure}[t]{0.49\textwidth}
        \centering
                \includegraphics[width=\textwidth]{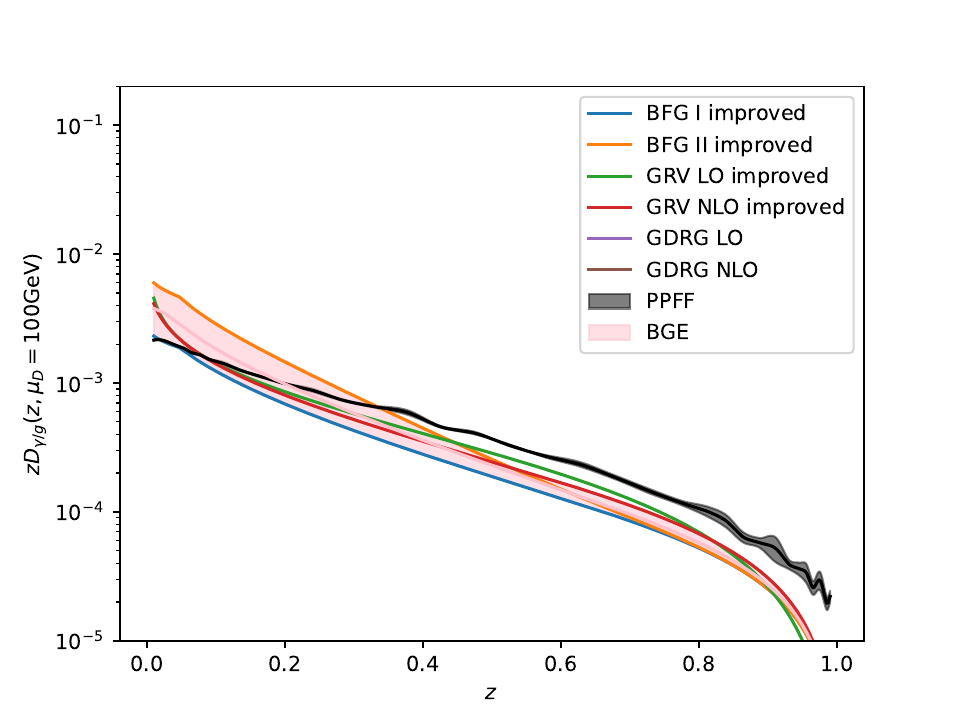}
                \caption{
                        Gluon.
                }
        \end{subfigure}
        \caption{
			FFs BFG~I, BFG~II, GRV~LO, GRV~NLO, GDRG~LO and GDRG~NLO  as well as the envelope set BGE and PPFF as a function of $z$ at $\mu_D = 100\,\mathrm{GeV}$ for the up quark, down quark, strange quark and gluon.
        }
        \label{fig:all_ff}
\end{figure}

\Cref{fig:all_ff} introduces a derived fragmentation-function set that we call \texttt{BFG\_GRV\_ENV} (BGE), constructed so that its three members span BFG~I, BFG~II, and GRV~NLO.
It is not meant as a statistically meaningful photon-FF uncertainty estimate, but rather as a practical demonstration of how heterogeneous photon-fragmentation-function variations can be packaged and propagated in \textsc{LHAPDF} in analogy with PDF error sets.
Because it combines parametrizations with different theoretical assumptions, it should be regarded as an illustrative reference-spread set rather than a FF uncertainty set for hadronic cross sections.
We opted to exclude GDRG from the envelope because it has no gluon contribution.
The central member is the arithmetic mean of these three sets, while the lower and upper members follow the point-by-point envelope hence we use the asymmetric \textsc{LHAPDF} grid \verb|ErrorType| \verb|hessian|.
In particular, the GRV sets are formulated in the $\mathrm{DIS}_\gamma$ scheme, while the BFG sets are given in the $\overline{\rm MS}$ scheme, and the various families also differ in their starting scales and non-perturbative inputs.
The conversion between the schemes is given in \cite{Gluck:1992zx} as
\begin{align}
        D^{\overline{\rm MS}}_{\gamma/q}(z,\mu_D) &= D^{\mathrm{DIS}_\gamma}_{\gamma/q}(z,\mu_D) - \frac{\alpha}{2\pi} e_q^2 \left[ \frac{1+(1-z)^2}{z} \log(z^2(1-z)) - 2\frac{1-z}{z} \right]\,, \\
        D^{\overline{\rm MS}}_{\gamma/g}(z,\mu_D) &= D^{\mathrm{DIS}_\gamma}_{\gamma/g}(z,\mu_D)\,,
\end{align}
where $\alpha= 1/137$ per \cite{Gluck:1991ee}.
Our purpose is therefore not to reinterpret these sets as a single theoretically uniform fit, but rather to provide them in a common \textsc{LHAPDF} representation and to use them as practical reference models for comparison with the \textsc{Pythia}-based extraction.
This produces a collider-oriented reference set that can be stored in \textsc{LHAPDF} alongside the original parameterizations and compared directly with the shower-extracted \textsc{Pythia} distributions tagged Pythia Photon Fragmentation Functions (\texttt{PPFF}) discussed in the previous section.
The resulting \texttt{PPFF} \textsc{LHAPDF} grid also includes one \verb|symmhessian| member for each flavour channel, which can be used to propagate the statistical uncertainty of the Monte-Carlo extraction to downstream observables.
\Cref{fig:all_ff} shows the original fragmentation functions as a function of $z$ at $\mu_D = 100\,\mathrm{GeV}$ for the up quark, down quark, strange quark and gluon.
We include the GDRG sets in this comparison for completeness, despite it not being part of the combined envelope set.
At a scale of 100 GeV we find that there is quite a spread between the different fragmentation functions, especially for the gluon.
The distinctive reduction at $z\approx 0.8$ of the GDRG~NLO quark distributions is also visible in Fig.~8 of Ref.~\cite{Gehrmann-DeRidder:1998bju}.
Encouragingly, we also find that the \textsc{Pythia} extraction is consistent with the spread of the different fragmentation functions, prompting confidence in the extraction method.
Unsurprisingly, the high $z$ region gluon contribution shows some small numerical fluctuations as captured in the uncertainty band of the \texttt{PPFF} set.

\section{Phenomenology}
Next we study the impact of different fragmentation functions on prompt photon production at NLO in QCD.
For all following computations we use \textsc{MCFM}'s prompt photon production at NLO with the \texttt{NNPDF40\_nlo\_as\_01180} PDF set \cite{NNPDF:2021njg}.
We use a modified version of \textsc{MCFM} with a new custom interface to query fragmentation functions from \textsc{LHAPDF}.\footnote{
The interface to \textsc{LHAPDF} for fragmentation functions can be found here: \url{https://gitlab.com/APN-Pucky/mcfm/-/tree/lhaff}
}\footnote{
A bug in MCFM's interpolation routines for BFG~I and BFG~II was found and fixed in this work.
It is to be included in future MCFM release.
Exactly this kind of consistency problems between fragmentation function providers can be avoided by using a standardized interpolator and library like \textsc{LHAPDF}.
}
In order to assess the uncertainty coming from the fragmentation functions we choose and vary the fragmentation scale $\mu_D=p_{T,\gamma}$ by factors of two up and down.

\subsection{Impact on non-isolated prompt photon production}

\begin{figure}
        \centering
        \begin{subfigure}[t]{0.49\textwidth}
        \centering
                \includegraphics[width=\textwidth]{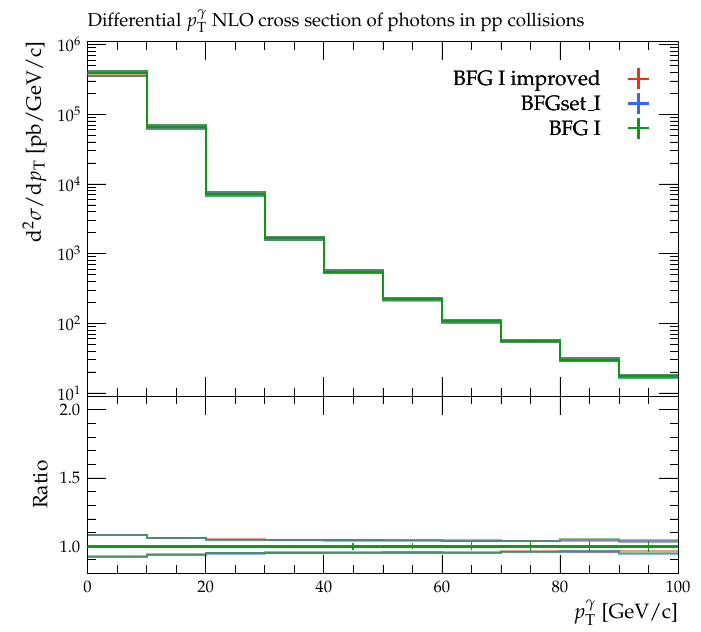}
                \caption{}\label{fig:mcfm_pheno:bfgs}
        \end{subfigure}
        \begin{subfigure}[t]{0.49\textwidth}
        \centering
                \includegraphics[width=\textwidth]{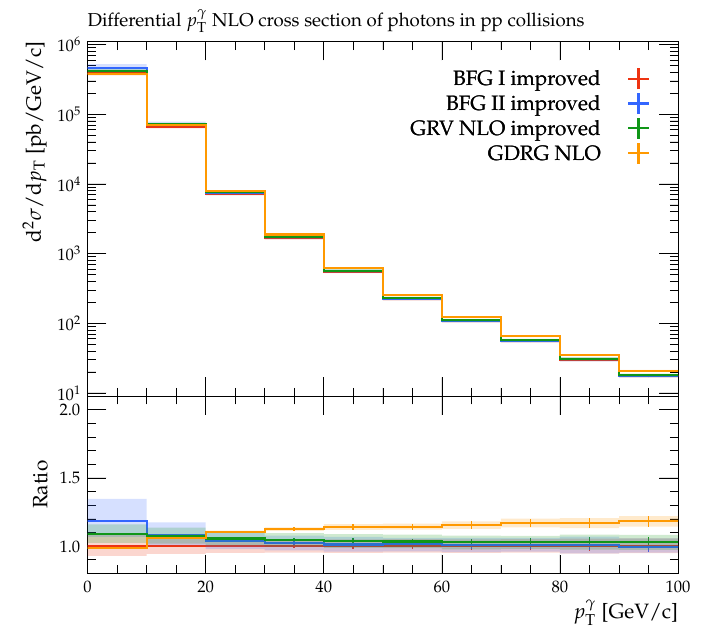}
                \caption{}\label{fig:mcfm_pheno:old}
        \end{subfigure}
        \begin{subfigure}[t]{0.49\textwidth}
        \centering
                \includegraphics[width=\textwidth]{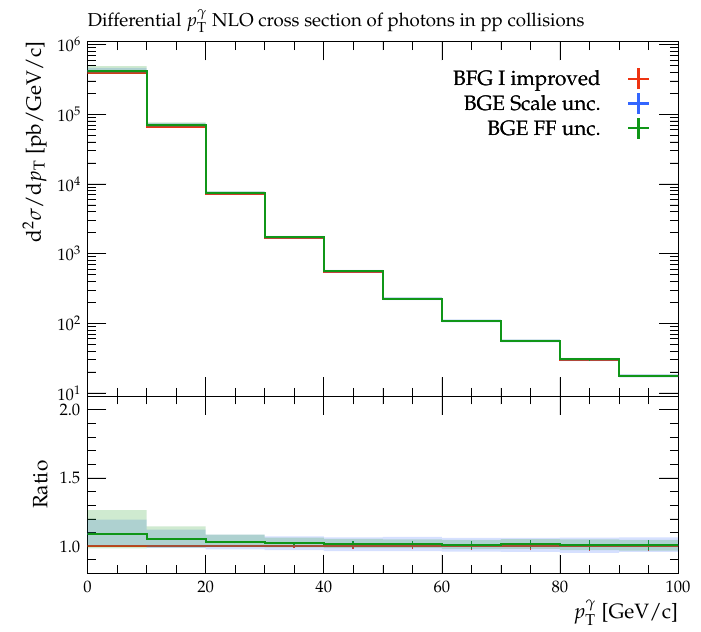}
                \caption{}\label{fig:mcfm_pheno:apn}
        \end{subfigure}
        \begin{subfigure}[t]{0.49\textwidth}
        \centering
                \includegraphics[width=\textwidth]{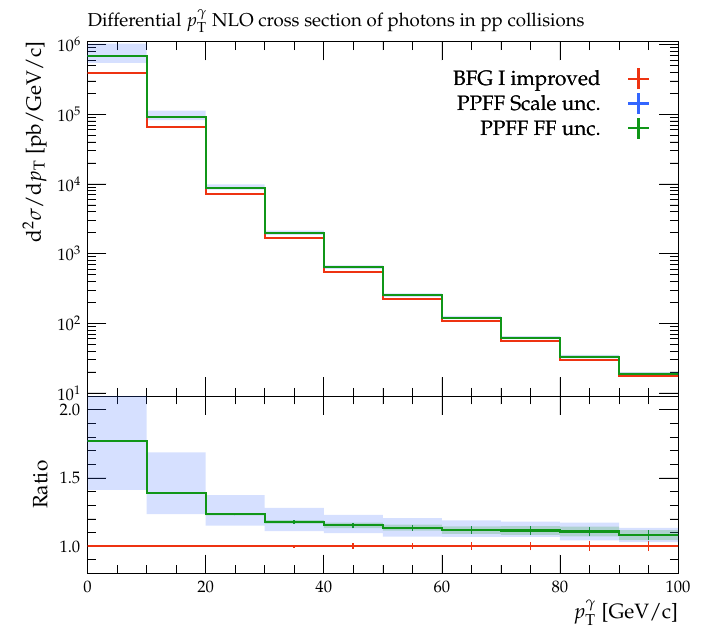}
                \caption{}\label{fig:mcfm_pheno:ppff}
        \end{subfigure}
        \caption{
                \textsc{MCFM} NLO prompt photon production with various fragmentation functions as a function of photon transverse momentum and rapidity.
                The uncertainty band is obtained by varying the fragmentation scale $\mu_D=p_{T,\gamma}$ by factors of two up and down.
                Note that, for visibility in the third plot only the outline of the uncertainty band is shown, since they are almost exactly equal for all three curves.
                The vertical bars indicate the statistical uncertainty of the Monte Carlo integration.
        }
        \label{fig:mcfm_pheno}
\end{figure}

\Cref{fig:mcfm_pheno} shows the impact of different fragmentation functions on the NLO prompt photon production cross section as a function of photon transverse momentum.
No isolation is applied to the photon, so the fragmentation contribution is not suppressed.
The variation of the scales is illustrated as the colored envelope band around the central value.
We first compare interpolation choices, then the envelope sets, and finally PPFF.

The \cref{fig:mcfm_pheno:bfgs} shows the comparison between the original \textsc{MCFM} BFGset I, the new \textsc{LHAPDF} linear interpolated BFG~I and BFG~I improved using logcubic interpolation as a function of photon transverse momentum.
We see that the results are consistent between the different interpolation methods up to very small numerical fluctuations.
In \cref{fig:mcfm_pheno:old} we find that for all fragmentation functions the uncertainty band grows up to $\pm10$\% to $\pm20$\% towards very low transverse momentum, except for GDRG~NLO, which has a very small uncertainty band compared to the remaining numerical uncertainties.
One way that experiments such as ALICE developed to avoid large fragmentation function uncertainties is to study ratios of prompt photon production in different collision systems, such as pp over pPb, which cancels out the fragmentation scale uncertainty \cite{ALICE:2025bnc}.
The resulting BGE envelope should be interpreted here as an illustrative model-spread band demonstrating the practical propagation of FF variations through \textsc{LHAPDF}, rather than as a rigorous theoretical uncertainty estimate.
Next \cref{fig:mcfm_pheno:apn} we also see that the uncertainty band of the fragmentation set variation ("BGE FF unc.") exceeds the uncertainty band of the fragmentation scale variation ("BGE scale unc.") for $p_T$ below 20 GeV.
In the ratio plot we also see that the BGE FF lower bound coincides with the BFG~I curve as is expected from the construction of the BGE set.
Finally, \cref{fig:mcfm_pheno:ppff} shows the \textsc{Pythia} extracted fragmentation functions with scale and fragmentation function uncertainty bands.
The fragmentation function uncertainty band is very small, since it is only the statistical uncertainty of the Monte Carlo extraction.

Two striking differences are that PPFF is significantly larger than the other fragmentation functions at transverse momentum $p_T$ below 20 GeV, whereas the GDRG~NLO set gives larger cross sections towards 100 GeV. 
This difference should be interpreted with caution since the PPFF extraction identifies the fragmentation scale with $p_{T,\text{parton}}$ whereas the fixed-order phenomenology uses $\mu_D=p_{T,\gamma}$.


\begin{figure}
        \centering
        \begin{subfigure}[t]{0.49\textwidth}
        \centering
                \includegraphics[width=\textwidth]{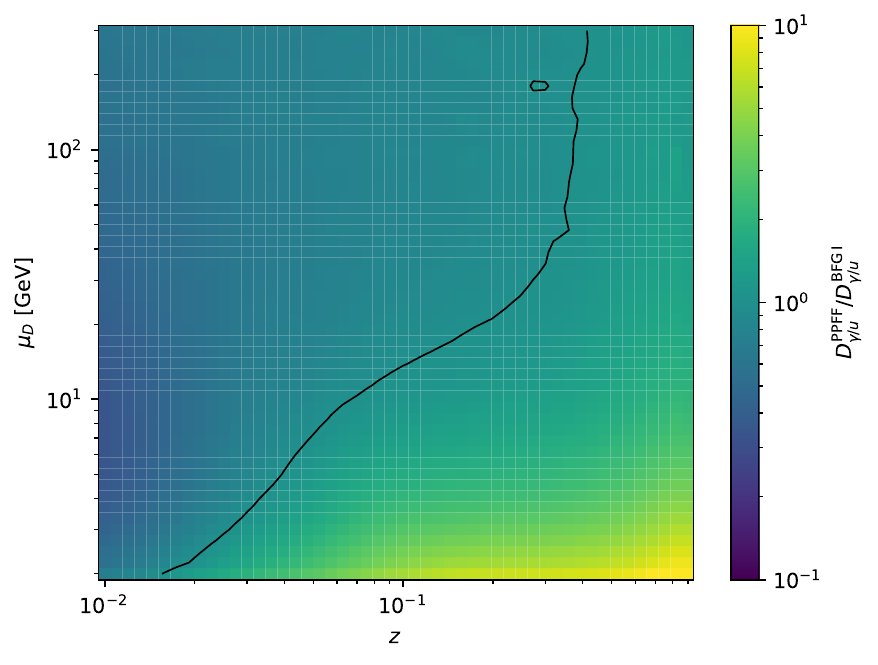}
        \end{subfigure}
        \begin{subfigure}[t]{0.49\textwidth}
        \centering
                \includegraphics[width=\textwidth]{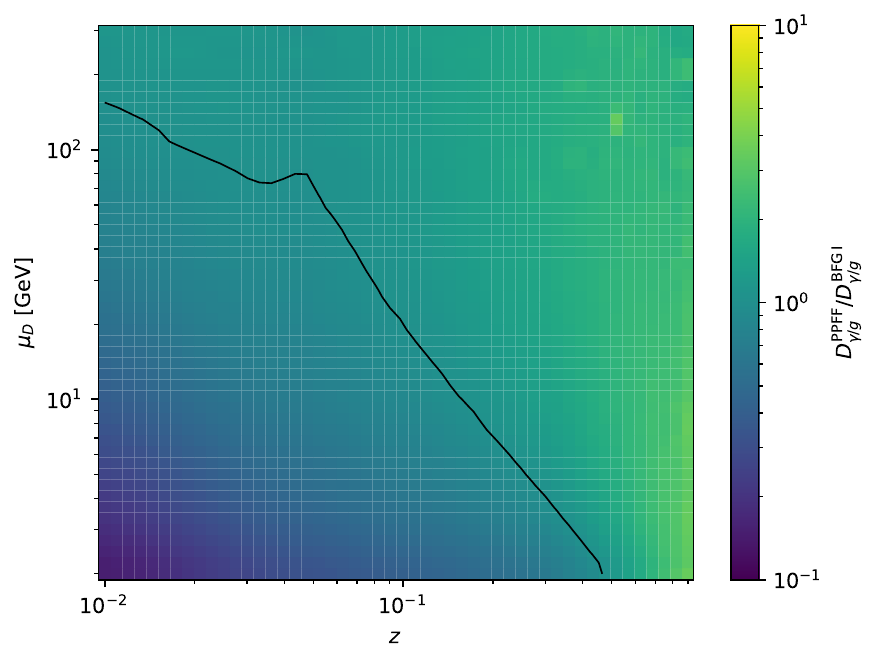}
        \end{subfigure}
        \caption{
                Ratios of the photon fragmentation functions extracted from \textsc{Pythia} (PPFF) to the BFG~I improved parametrization as functions of the photon momentum fraction $z$ and fragmentation scale $\mu_D$, shown for up-quark (left) and gluon (right) fragmentation.
                The black contour indicates equal fragmentation functions, $D_{\gamma/i}^{\mathrm{PPFF}}/D_{\gamma/i}^{\mathrm{BFG,I}}=1$.
                The comparison highlights the substantial differences between the parametrizations at low fragmentation scales, which contribute to the enhanced sensitivity to the choice of fragmentation function observed for non-isolated photons.
        }
        \label{fig:heat}
\end{figure}

In order to understand why the PPFF is significantly larger than the other fragmentation functions at transverse momentum $p_T$ below 20 GeV, we show in \cref{fig:heat} the ratio of the PPFF to the BFG~I improved as functions of the photon momentum fraction $z$ and fragmentation scale $\mu_D$, shown for up-quark and gluon fragmentation.
Other quark flavours show similar behaviour to the up-quark and are not shown here.
The largest differences are in the quark channels at low fragmentation scales and medium to large $z$ values, which is exactly the region that contributes to the enhanced sensitivity to the choice of fragmentation function observed for non-isolated photons.
The gluon channel shows a different behaviour, where the PPFF is smaller by a factor of 10 when both $z$ and the fragmentation scales are low.
The absence of an explicit non-perturbative boundary condition in PPFF is expected to be most relevant at low scales, while at larger $\mu_D$ the photon FF becomes increasingly dominated by perturbative evolution, $D_{\gamma/q}^{\mathrm{pert}}\sim \alpha P_{\gamma q}\log(\mu_D^2/\mu_0^2)$.
This naturally leads to larger differences between the \texttt{PPFF} and parameterizations with an explicit non-perturbative input at low $\mu_D$.

\begin{figure}
        \centering
        \begin{subfigure}[t]{0.49\textwidth}
        \centering
                \includegraphics[width=\textwidth]{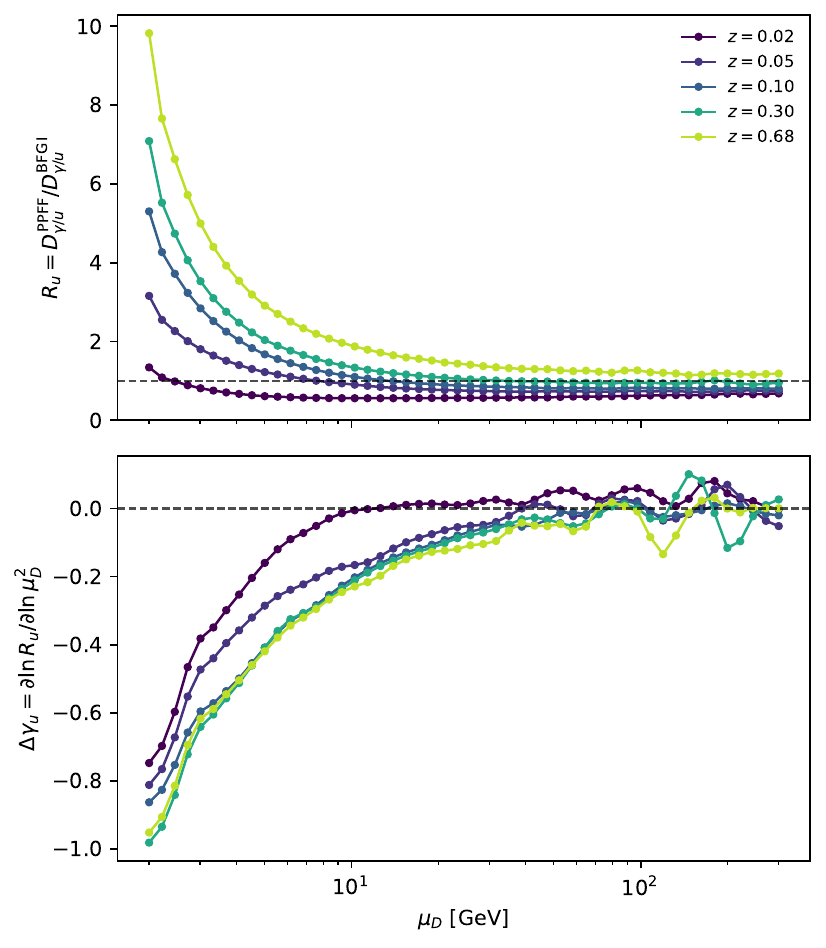}
        \end{subfigure}
        \begin{subfigure}[t]{0.49\textwidth}
        \centering
                \includegraphics[width=\textwidth]{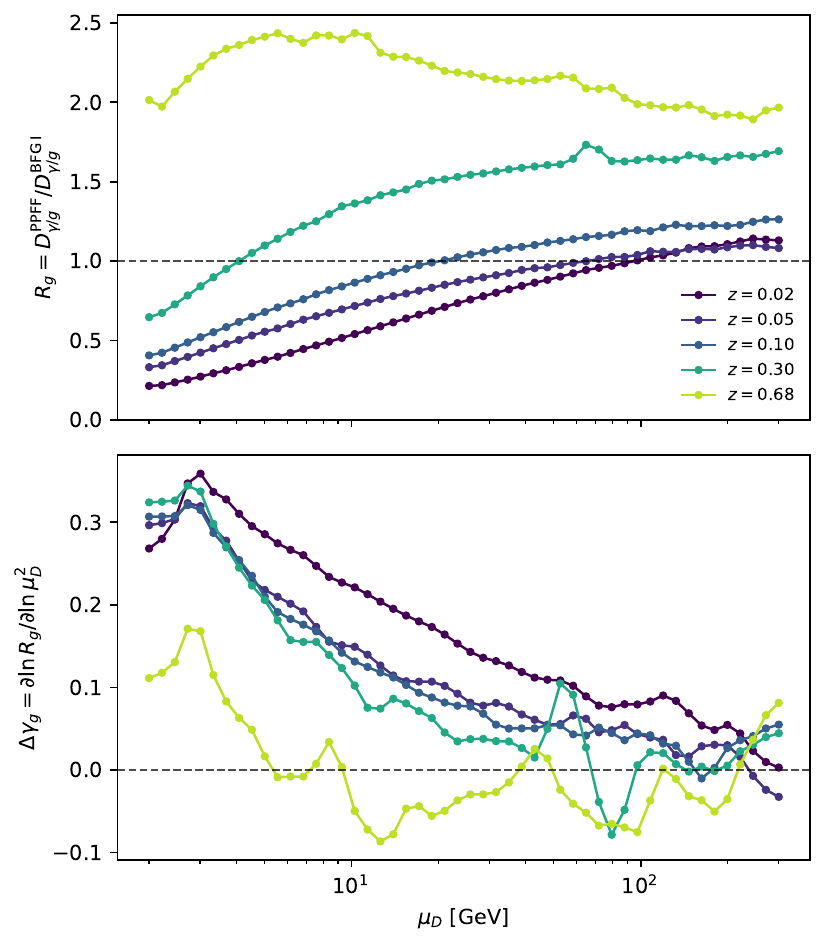}
        \end{subfigure}
        \caption{
                Fixed-$z$ slices of the same ratio $R_i=D_{\gamma/i}^{\mathrm{PPFF}}/D_{\gamma/i}^{\mathrm{BFG,I}}$ shown in \cref{fig:heat}, plotted as a function of $\mu_D$ for several values of $z$, for the up-quark (left) and gluon (right) channels (upper panels).
                The dashed horizontal line there marks $R_i=1$.
                The lower panels show the corresponding difference of logarithmic scale derivatives $\Delta\gamma_i^\text{eff}$, defined in \cref{eq:deltagamma}, for the same $z$ slices.
                The dashed horizontal line there marks $\Delta\gamma_i^\text{eff}=0$, i.e.\ an equal logarithmic scale dependence of the \texttt{PPFF} and BFG~I fragmentation functions.
        }
        \label{fig:heat_slice}
\end{figure}

The same ratio underlying \cref{fig:heat} is shown as fixed-$z$ slices in the upper panels of \cref{fig:heat_slice}, allowing this trend to be quantified precisely.
At the lowest scale shown the quark ratio $R_i = D_{\gamma/i}^{\mathrm{PPFF}}/D_{\gamma/i}^{\mathrm{BFG,I}}$ ranges from about $0.7$ at small $z$ up to $10$--$15$ at large $z$.
It then essentially decreases monotonically with $\mu_D$ at every fixed $z$, reaching a plateau within about $30\%$ of unity already by $\mu_D$ of a few tens of GeV.
The gluon channel shows a different trend.
Initially, the ratio increases with $\mu_D$ at fixed $z$, from about $0.25$ at small $z$ to a plateau of $R_g \sim 1\text{--}2$.
The two channels, however, reach their respective plateaux at a different pace: the gluon-channel ratio keeps drifting slowly over almost the full range shown before flattening out only towards $\mu_D\sim100\,\mathrm{GeV}$.
The quark channel receives its point-like contribution already at order $\alpha$, present from the lowest scales, whereas the gluon channel has no leading-order point-like term, $P_{\gamma g}(z)=0$, and only arises at the higher order $\mathcal{O}(\alpha\alpha_s)$ through QCD mixing with the quark channel.
It therefore requires a correspondingly longer evolution in $\mu_D$ for \texttt{PPFF} and BFG~I to settle into a common asymptotic behaviour.
This illustrates the heightened importance of the non-perturbative input for the gluon channel and is itself consistent with the structure of \cref{eq:evoff}.
We stress, however, that this plateau value does not have to be exactly unity.
At leading order the splitting kernel $P_{\gamma q}(z)$ entering \cref{eq:evoff} is universal, so the quark channels of \texttt{PPFF} and BFG~I would only be expected to approach the same value logarithmically slowly as $\mu_D\to\infty$.
However, BFG~I includes explicit factorization-scheme dependent beyond-leading-logarithm (BLL) corrections to the splitting kernels and is defined in a specific factorization scheme \cite{Bourhis:1997yu}.
The \textsc{Pythia} shower, in contrast, generates photon radiation through leading-logarithmic, leading-color splitting probabilities ordered in its own evolution variable and does not correspond to any standard factorization scheme.
These order- and scheme-related differences are expected to leave a residual offset, so an exact convergence of the ratio to unity is not what we expect.
The stability of the plateau, rather than its absolute value, is what we regard as meaningful.
To compare the logarithmic scale dependence we define
\begin{equation}
\label{eq:deltagamma}
\Delta \gamma_i^\text{eff} = \frac{\partial \log D_{\gamma/i}^{\mathrm{PPFF}} }{\partial \log \mu_D^2} -  \frac{\partial \log D_{\gamma/i}^{\mathrm{BFG~I}} }{\partial \log \mu_D^2} = \frac{\partial \log R_i}{\partial \log \mu_D^2}\,,
\end{equation}
where we approximate the derivative by finite differences between the $\mu_D$ values shown.
If $\Delta \gamma_i^\text{eff} \to 0$, both FFs show the same scale dependence, even as $R_i\neq 1$.
In the lower panels of \cref{fig:heat_slice}, this is observed for both the quark and gluon channels, at different convergence rates, confirming that the scale dependence of the \textsc{Pythia} extraction is consistent with the expected perturbative evolution of the photon fragmentation functions.
A more direct check of the analytic evolution of \texttt{PPFF} compared against the extraction at different scales is left for future work, once a public implementation of the evolution equations is available.


\subsection{Comparison with LHC 7 TeV isolated prompt photon measurements}

In order to compare against isolated prompt photon measurements we use \textsc{Rivet} 4.1.3 and \textsc{MCFM} 10.2 with the interface between them developed in \cite{Jezo:2024wsc}.\footnote{
The interface has been updated to support Rivet 4 here: \url{https://gitlab.com/APN-Pucky/mcfm/-/tree/rivet4}
}
In particular, we look at ALICE~\cite{ALICE:2019rtd} and ATLAS~\cite{Aad:2011tw} analyses.\footnote{
We also looked at CMS~\cite{Chatrchyan:2013mwa}, but it seemed like the corresponding unvalidated \textsc{Rivet} analysis does not include the proper isolation prescription to match the experimental data.
} 
The published analysis isolation is applied by the subsequent \textsc{Rivet} analysis while a looser isolation is imposed in \textsc{MCFM} as a generation-level preselection to improve Monte-Carlo efficiency.
We abstain from visualizing the fragmentation function uncertainty band in the following plots, since it is negligible after isolation.

\begin{figure}
        \centering
        \begin{subfigure}[t]{0.49\textwidth}
        \centering
                \includegraphics[width=\textwidth]{
                        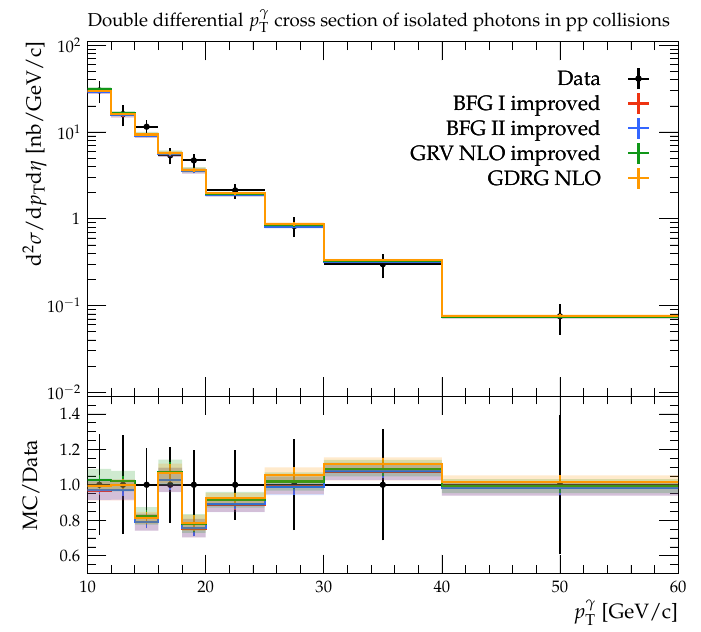
                }
        \end{subfigure}
        \begin{subfigure}[t]{0.49\textwidth}
        \centering
                \includegraphics[width=\textwidth]{
                        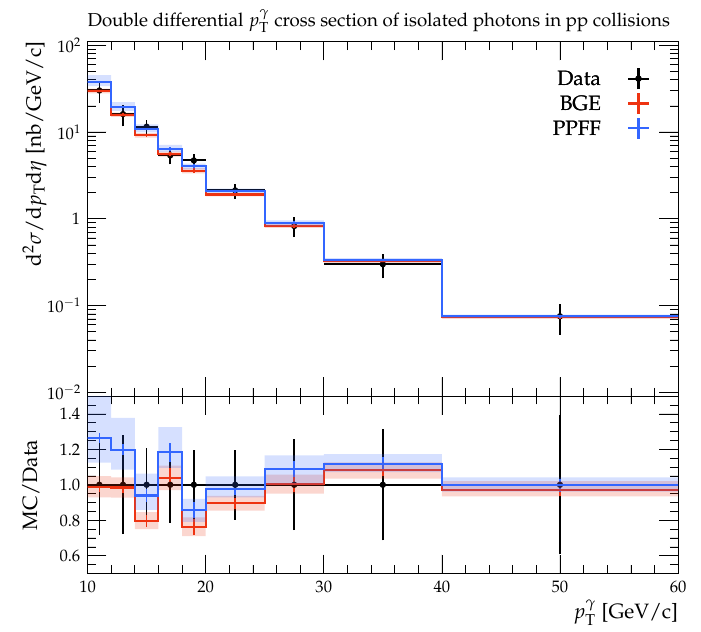
                }
        \end{subfigure}
        \caption{
                ALICE central ($|\eta| < 0.27$) isolated prompt photon production at 7 TeV compared to various \textsc{MCFM} NLO predictions with the different fragmentation functions as a function of photon transverse momentum.
                The vertical bars indicate the statistical uncertainty of the Monte Carlo integration and the band represents the scale variation uncertainty.
        }
        \label{fig:alice}
\end{figure}

\Cref{fig:alice} shows the comparison between ALICE isolated prompt photon production at 7 TeV and \textsc{MCFM} NLO predictions with the different fragmentation functions as a function of photon transverse momentum.
While the PPFF is again significantly larger at transverse momentum $p_T$ below 15 GeV and exceeds experimental data in those bins it is also in better agreement with the data from 15 GeV to 25 GeV than the other fragmentation functions.
All fragmentation functions are in 1 $\sigma$ agreement with the experimental data, which has large uncertainties of about $\pm20\%$.

\begin{figure}
        \centering
        \begin{subfigure}[t]{0.49\textwidth}
        \centering
                \includegraphics[width=\textwidth]{
                        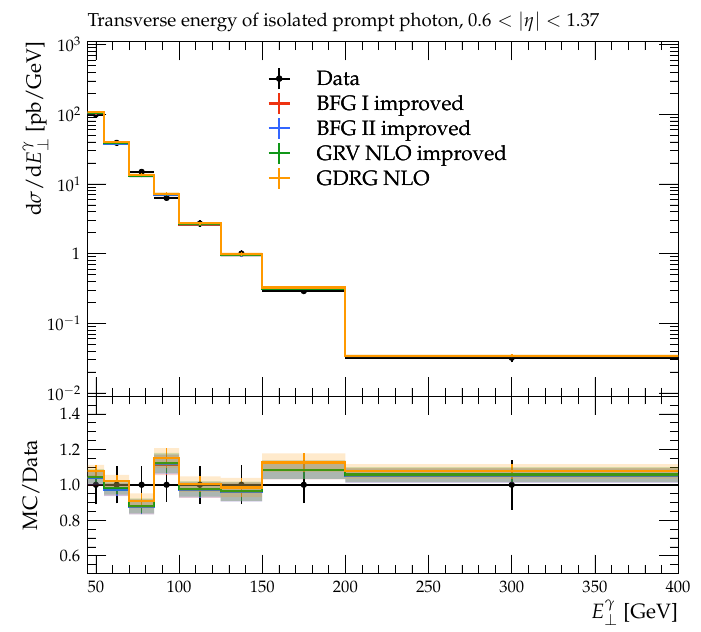
                }
        \end{subfigure}
        \begin{subfigure}[t]{0.49\textwidth}
        \centering
                \includegraphics[width=\textwidth]{
                        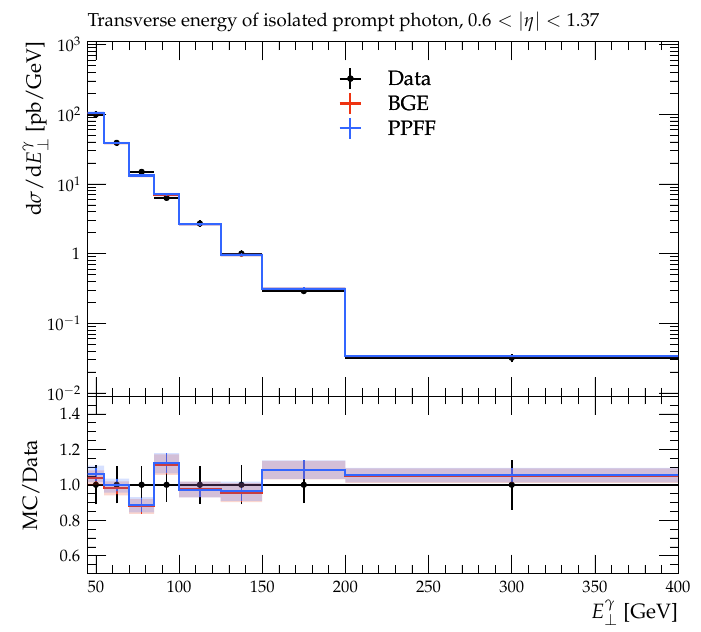
                }
        \end{subfigure}
        \caption{
                ATLAS transverse momentum at intermediate rapidity of isolated prompt photons at 7 TeV compared to MCFM NLO predictions with the various fragmentation functions.
                The vertical bars indicate the statistical uncertainty of the Monte Carlo integration and the band represents the scale variation uncertainty.
        }
        \label{fig:atlas}
\end{figure}

\Cref{fig:atlas} shows the comparison between ATLAS isolated prompt photon production at 7 TeV and \textsc{MCFM} NLO predictions with the different fragmentation functions as a function of photon transverse momentum above 50 GeV in a less central rapidity regime.
The previously larger high transverse momentum $p_T$ GDRG~NLO FF gives slightly larger predictions which are however negligible compared to experimental and scale variation uncertainties.
Further, there is no visible difference between the BFG~I, BFG~II, GRV~NLO, BGE and PPFF, which means that the isolation removes most of the fragmentation function contribution.
Overall, the dominant direct NLO prompt photon contribution is in very good agreement with the experimental data.

\section{Conclusion and Outlook}

We have demonstrated how the existing capabilities of \textsc{LHAPDF} can be utilized to store and interpolate photon fragmentation functions in a portable grid format.
Since existing tools such as \textsc{JetPhox}, \textsc{MCFM}, and \textsc{NNLOJet} already support \textsc{LHAPDF} for parton distribution functions, this opens the door to a more flexible and consistent treatment of photon fragmentation functions in phenomenological studies.
Further uncertainty treatment is arguably more advanced in PDF fits than in photon fragmentation functions, so this approach also provides a convenient framework for future global fits of photon fragmentation functions.
In \cref{tab:sets} we summarize the photon fragmentation function sets added to \textsc{LHAPDF} as part of this publication.
We found the ALICE and ATLAS isolated prompt-photon measurements at 7 TeV to be in very good agreement, within uncertainties, with the NLO \textsc{MCFM} predictions obtained with the different fragmentation functions, confirming that the isolation strongly suppresses the residual fragmentation sensitivity.

Future prompt-photon measurements with less stringent isolation criteria, or extended to lower photon transverse momentum, would enhance sensitivity to fragmentation contributions and could help discriminate between different photon-fragmentation models.
It would be valuable to compare the default \textsc{Pythia} shower with alternative shower formalisms such as \textsc{Vincia}~\cite{Brooks:2020upa} and to repeat the extraction with NLO-matched events from \textsc{POWHEG-BOX}, which would incorporate higher-order hard-radiation effects before the photon fragmentation analysis.
On the phenomenology side, it would also be useful to compare the parton-based extraction adopted here with a jet-based definition with hadronization effects that is closer to experimentally accessible photon-in-jet observables.
On the evolution side, extending public libraries such as \textsc{EKO}~\cite{Candido:2022tld,barontini_2026_21486740} or \textsc{APFEL++}~\cite{Bertone:2017gds,valerio_bertone_2024_10522420} with the QED ingredients needed for photon-fragmentation evolution would make it possible to evolve the shower-extracted low-scale distributions to higher scales in a more systematic way.
The recent developments in $\gamma$\textsc{EKO}~\cite{Chithirasreemadam:2026mqp} and \textsc{APFEL++} are promising starting points for this.
This would also serve as a consistency check of the shower evolution implementation.
Together, these developments would help establish photon fragmentation functions as a quantitatively constrained ingredient of precision prompt-photon phenomenology.

\begin{table}[t]
        \centering
        \caption{
                Summary of the photon-fragmentation-function sets provided in \textsc{LHAPDF}.
                The BGE set contains three members corresponding to the lower envelope, central value, and upper envelope, and is intended as a demonstrator for FF systematic variations and reference spreads rather than as a fitted uncertainty estimate.
        }
        \label{tab:sets}
        \begin{tabular}{llcc}
                \toprule
                Interpolator & Set name & Scheme &Members \\
                \midrule
                linear   & \texttt{BFG\_I\_FF\_photon} & $\overline{\rm MS}$ & 1 \\
                logcubic & \texttt{BFG\_I\_improved\_FF\_photon} & $\overline{\rm MS}$ & 1 \\
                linear   & \texttt{BFG\_II\_FF\_photon} & $\overline{\rm MS}$ & 1 \\
                logcubic & \texttt{BFG\_II\_improved\_FF\_photon} & $\overline{\rm MS}$ & 1 \\
                linear   & \texttt{GRV\_LO\_FF\_photon} & $\overline{\rm MS}$ & 1 \\
                logcubic & \texttt{GRV\_LO\_improved\_FF\_photon} & $\overline{\rm MS}$ & 1 \\
                linear   & \texttt{GRV\_NLO\_FF\_photon}  & $\overline{\rm MS}$& 1 \\
                logcubic & \texttt{GRV\_NLO\_improved\_FF\_photon} & $\overline{\rm MS}$ & 1 \\
                logcubic & \texttt{GDRG\_LO\_FF\_photon} & $\overline{\rm MS}$  & 1 \\
                logcubic & \texttt{GDRG\_NLO\_FF\_photon} & $\overline{\rm MS}$  & 1 \\
                logcubic & \texttt{BFG\_GRV\_ENV\_FF\_photon} & $\overline{\rm MS}$ & 3 \\
                logcubic & \texttt{Pythia\_Pert\_FF\_photon} & - & 2 \\
                \bottomrule
        \end{tabular}
\end{table}

\backmatter

\bmhead{Acknowledgements}

Part of the data was produced in parallel using \textsc{GNU Parallel} \cite{tange_2025_16289600} and analysed in \textsc{babyyoda} \cite{alexander_puck_neuwirth_2026_19353784}.
Calculations for this publication were performed on the HPC cluster PALMA II of the University of Münster, subsidized by the DFG (INST 211/667-1).

\begin{appendices}

\section{Combining transverse momentum slices} \label{sec:ptslices}

\begin{figure}
        \centering
        \begin{subfigure}[t]{0.49\textwidth}
                \centering
                \includegraphics[width=\linewidth]{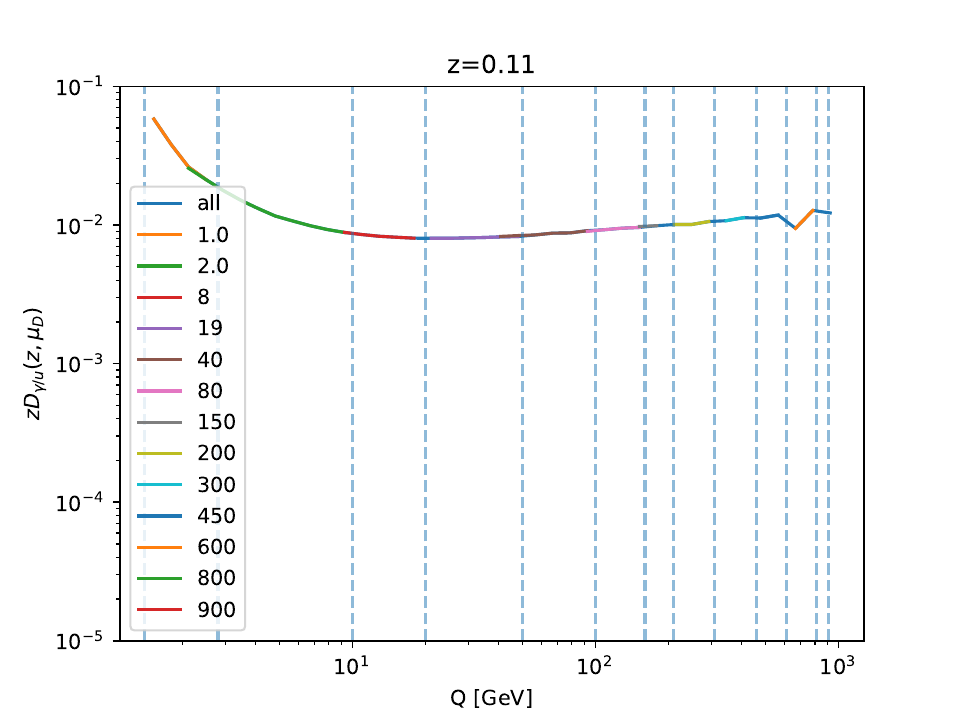}
                \caption{up, $z = 0.11$.}
        \end{subfigure}
        \begin{subfigure}[t]{0.49\textwidth}
                \centering
                \includegraphics[width=\linewidth]{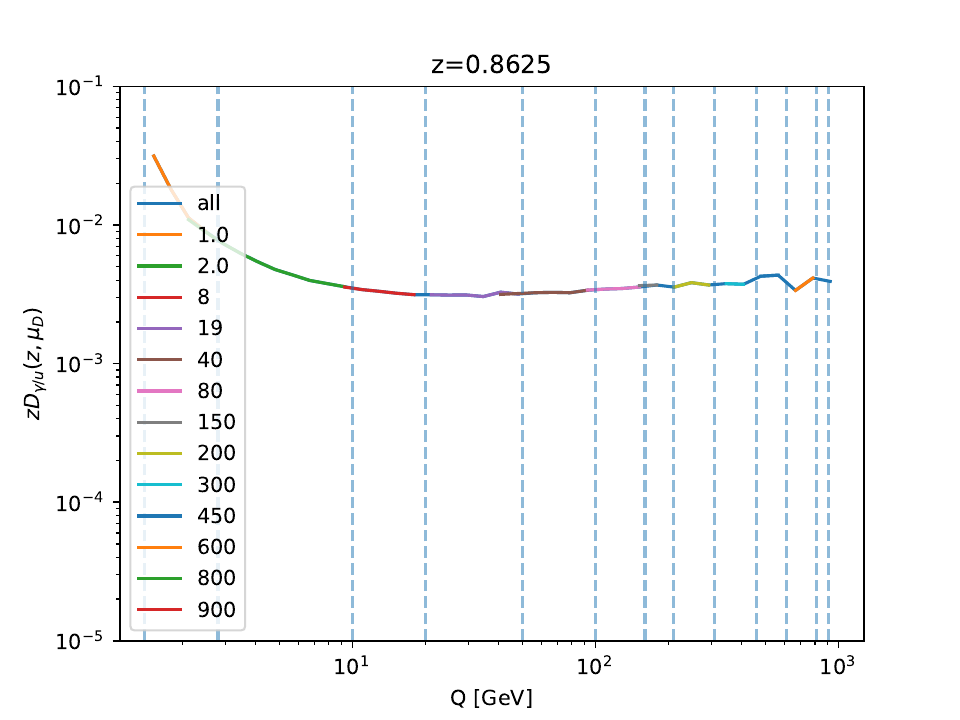}
                \caption{up, $z = 0.86$.}
        \end{subfigure}
        \begin{subfigure}[t]{0.49\textwidth}
                \centering
                \includegraphics[width=\linewidth]{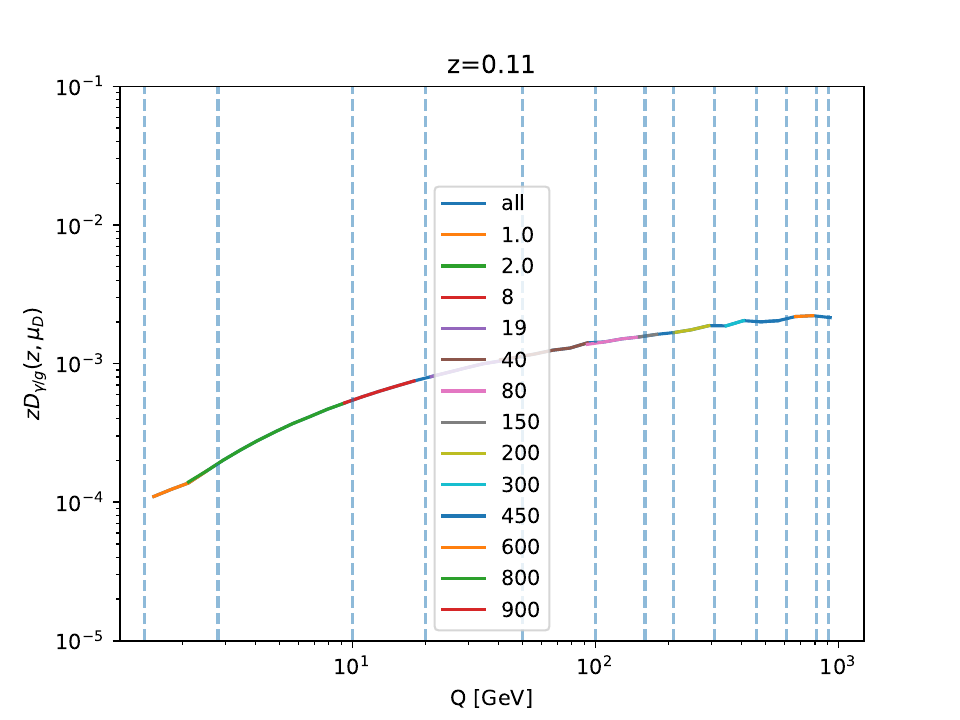}
                \caption{gluon, $z = 0.11$.}
        \end{subfigure}
        \begin{subfigure}[t]{0.49\textwidth}
                \centering
                \includegraphics[width=\linewidth]{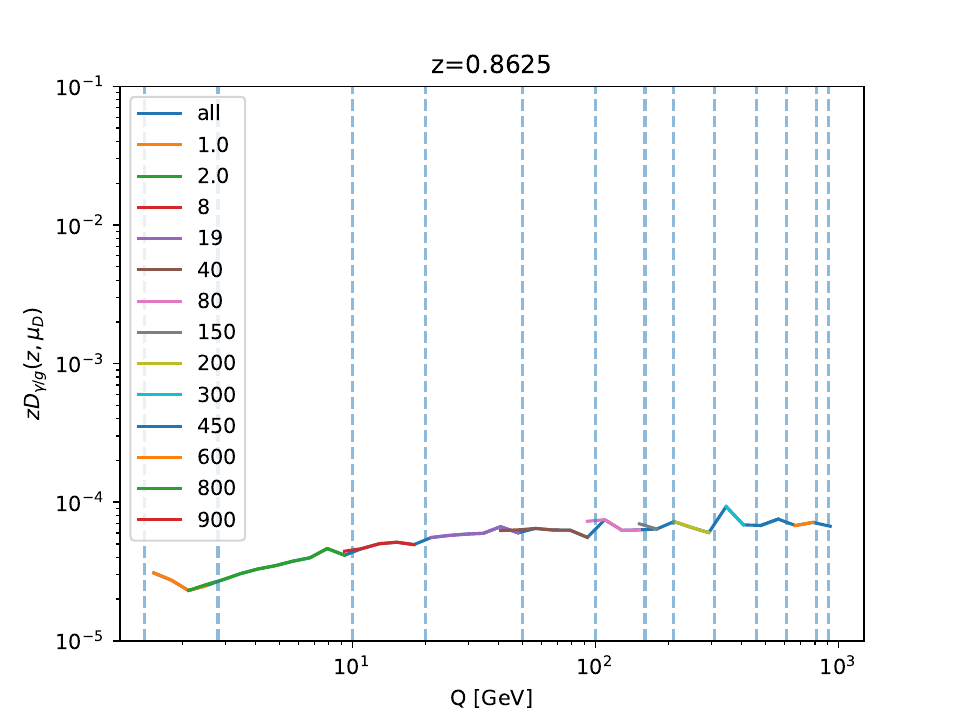}
                \caption{gluon, $z = 0.86$.}
        \end{subfigure}
        \caption{
                Multiple \textsc{Pythia} runs with different hard-process transverse-momentum slices are combined to populate full $(z,\mu_D)$ plane.
                }
        \label{fig:pythia_ptslices}
\end{figure}

Since it is difficult to populate the full $(z,\mu_D)$ plane with a single \textsc{Pythia} run, we combine multiple runs with different hard-process transverse-momentum slices.
The process of stitching together the different slices is illustrated in \cref{fig:pythia_ptslices}, where each curve indicates a \verb|PhaseSpace:pTHatMin| setting in \textsc{Pythia} and the vertical dashed lines indicate the boundaries of the transverse-momentum slices.

\section{Python package: \textsc{lhagrid}}
\label{lhagrid}

For the creation of the \textsc{LHAPDF} grids, we developed a minimal Python package called \textsc{lhagrid}, which is available on GitLab at \url{https://gitlab.com/APN-Pucky/lhagrid}.
It provides a simple interface for reading and writing \textsc{LHAPDF}-compatible grids, as well as metadata stored therein.
The package is available on PyPI and can be installed via \texttt{pip install lhagrid}.
The source code is licensed under the GPLv3 License, and contributions are welcome.

\section{Code availability} \label{sec:code_availability}
The grid sampling of the photon fragmentation functions is given in the respective folders of \url{https://gitlab.com/APN-Pucky/lhaff}.
The conversion to \textsc{LHAPDF} grids is performed in \verb|lhaff.ipynb|, which is available in the same repository.
The \textsc{Rivet} analysis used to extract the \textsc{Pythia} photon fragmentation functions is available at \url{https://gitlab.com/APN-Pucky/rivet-MC_PHOTONFRAGMENTATION}.
In the same repository, the \textsc{Pythia} configuration files are available in the \verb|pythia| folder.
The modified MCFM code and \verb|input.ini| cards can be found under \url{https://gitlab.com/APN-Pucky/mcfm/-/tree/lhaff}.




\end{appendices}


\bibliography{sn-bibliography}

\end{document}